\documentstyle[prb,aps,amsfonts,amssymb,floats,epsfig]{revtex}
\begin{document}
\draft 
\twocolumn[\hsize\textwidth\columnwidth\hsize\csname @twocolumnfalse\endcsname
\title{Optical spectroscopy of pure and doped CuGeO$_3$}
\author{A. Damascelli\cite{byline} and D. van der Marel}
\address{Solid State Physics Laboratory, University of Groningen, Nijenborgh 4, 9747 AG Groningen, The Netherlands}
\author{G. Dhalenne and A. Revcolevschi} 
\address{Laboratoire de Chimie des Solides, Universit$\acute{e}$ de Paris-sud, B$\hat{a}$timent 414, F-91405 Orsay, France}
\date{March 19, 1999}
\maketitle
\begin{abstract}   
We investigated in detail the optical properties of several 
Cu$_{1-\delta}$Mg$_{\delta}$GeO$_3$  (with $\delta$\,=\,0,\,0.01), and  
CuGe$_{1-x}B_{x}$O$_3$ with {\em B}=Si ({\em x}\,=\,0,\,0.007,\,0.05,\,0.1), and Al 
($x$\,=\,0,\,0.01) single crystals, in the frequency range 20\,-\,32\,000 cm$^{-1}$. 
We report temperature dependent reflectivity and transmission measurements, performed with 
polarized light in order to probe the anisotropy of the crystals along the {\em b} and  
{\em c} axes, and optical conductivity spectra obtained by Kramers-Kronig transformation or  
direct inversion of the Fresnel formula. Special emphasis is given to the far-infrared 
phonon spectra. The temperature dependence of the phonon parameters is presented and discussed 
in relation to the soft mode issue in CuGeO$_3$. For $T\!<\!T_{\rm{SP}}$ we could detect zone 
boundary folded modes activated by the spin-Peierls phase 
transition. Following the temperature dependence of these modes, which shows the second order 
character of the phase transition, we were able to study the effect of doping on $T_{\rm{SP}}$. 
Moreover, in transmission experiments we detected a direct singlet-triplet excitation at 44 
cm$^{-1}$, across the magnetic gap, which is not understandable on the basis of the 
magnetic excitation spectrum so far assumed for CuGeO$_3$.  The optical activity of this 
excitation and its polarization dependence confirm the existence of a second (optical) magnetic 
branch, recently suggested on the basis of inelastic neutron scattering data. The anisotropy in 
the magnetic exchange constants along the $b$ axis, necessary for the optical triplet mode to 
gain a finite intensity, and the strong effect of Si substitution on the phonon spectra are 
discussed in relation to the alternative space group $P2_12_12_1$, recently proposed for 
CuGeO$_3$ in the high temperature uniform phase. 
\end{abstract}
%
\vskip2pc]
\narrowtext

\section{Introduction}

In 1993  Hase {\em et al.}\cite{hase} concluded, on the basis of magnetic susceptibility 
measurements, that  CuGeO$_3$ is showing a spin-Peierls 
(SP) phase transition, i.e., a  lattice distortion (due to the magneto-elastic 
coupling between the one-dimensional spin system and the three-dimensional phonon system) 
that occurs together with the formation of a spin-singlet ground state and the opening of a 
finite energy gap in the magnetic excitation spectrum. This 
magneto-elastic transition is driven by the magnetic 
energy gain due to dimerization of the antiferromagnetic (AF) exchange between the 
spin 1/2 moments of the Cu$^{2+}$ ions [arranged in weakly coupled one-dimensional (1D) 
CuO$_2$ chains in this material\cite{vollenkle}], which overcompensates 
the elastic energy loss resulting from the deformation of the lattice.\cite{pytte,crossfi} 
In the SP ordered phase, the Cu$^{2+}$ magnetic moments form singlet dimers along the chains 
and spin triplet excitations are gapped.\cite{hase}
 
The SP nature of the phase transition  in  CuGeO$_3$ was inferred from the isotropic drop in 
the magnetic susceptibility at the transition temperature $T_{\rm{SP}}$=14 K,  and from the 
reduction of $T_{\rm{SP}}$  upon increasing the intensity of an applied magnetic field,\cite{hase} 
as theoretically expected for SP systems.\cite{pytte,crossfi,bulaevskii,cross} This initial claim 
was later confirmed by an impressive variety of experimental results.\cite{boucher} The gap in the 
magnetic excitation spectrum was directly observed with inelastic neutron 
scattering,\cite{nishi} and the singlet-triplet nature of the gap was established with the same 
technique under application of a magnetic field: A splitting of the single gap 
into three distinct excitation branches was clearly detected.\cite{fujita} The dimerization 
of the Cu$^{2+}$ ions was observed and the lattice distortion very carefully investigated with 
both neutron and x-ray scattering.\cite{hirota,pouget,braden} Finally, the  phase transition 
from the dimerized to the incommensurate phase, expected in magnetic fields higher than a certain 
critical value,\cite{bulaevskii,cross} was found with field-dependent magnetization measurements 
at $H_c\!\approx\!12$ T,\cite{hase1} and the {\em H-T} phase diagram was studied in 
great detail.\cite{lorenz}

The discovery of the SP phase transition in  CuGeO$_3$ has renewed the interest in 
this phenomenon, observed previously in organic materials in the 1970's,\cite{bray,jacobs,huizinga}  
because the availability of large high-quality single crystals of pure and doped 
 CuGeO$_3$  made it possible to investigate this magneto-elastic transition by a very broad variety 
of experimental techniques.  This way, the traditional SP theory 
based on 1D AF chains with only nearest-neighbor (nn) magnetic couplings and a mean-field treatment of the 
3D phonon system,\cite{pytte,crossfi,bulaevskii} could be tested in all its expectations 
for a deeper understanding of the problem. 

Also optical techniques, like Raman and infrared spectroscopy, are very useful in investigating 
magnetic and/or structural phase transitions. Information on the nature of the electronic 
(magnetic) ground state, lattice distortion and interplay of electronic (magnetic) 
and lattice degrees of freedom can be obtained studying in detail the electronic (magnetic) 
excitations and the phonon spectrum, as a function of temperature. 
The aim of our experimental investigation of  CuGeO$_3$ with optical spectroscopy was, of course, 
to detect all possible `infrared signatures' of the SP phase transition in this material. 
However, we were in particular interested in a number of issues which could help us in 
understanding how far the classical SP picture\cite{pytte,crossfi} is appropriate for the 
case of CuGeO$_3$, namely: 

(\/{\em i}\,) Analysis of the phonon spectra in order to study the lattice distortion and verify 
the proposed crystal structures for both the high and the low temperature phase. 

(\/{\em ii}\,) Verify the presence of a soft mode in the phonon spectra, upon  passing through 
the SP transition. In fact, a well-defined soft mode  is expected in those  theoretical  models 
describing a SP system in terms of a linear coupling between lattice and magnetic  
degrees of freedom.\cite{bulaevskii,crossfi} 

(\/{\em iii}\,) Detect possible magnetic bound states and/or a magnetic continuum in the excitation 
spectrum which could tell us about the symmetry and the order (nnn versus nn) of the  
magnetic interactions. 

(\/{\em iv}\,) Study the influence of doping on the vibrational and electronic 
properties. 

Before proceeding to the experimental results, we will present in the 
next section the outcome of a group theoretical analysis, which will be 
useful in the discussion of the 
phonon spectra. In fact, the number and the symmetry of the infrared and Raman active phonons 
expected for the proposed high-temperature undistorted phase\cite{vollenkle} and the 
low-temperature SP phase\cite{hirota,braden} of CuGeO$_3$ can be obtained from a group 
theoretical analysis of the lattice vibrational modes.\cite{ascona} 
The results of this calculation will later be compared to the experimental data.
\begin{figure}[t]
\centerline{\epsfig{figure=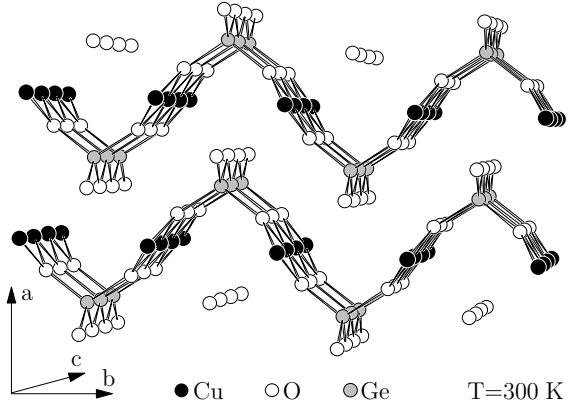,width=7.5cm,clip=}}
\vspace{.26cm}
\caption{Crystal structure of  CuGeO$_3$ in the high temperature (T=300 K) undistorted phase.}
\label{cgstru1}
\end{figure}

\section {Group Theoretical Analysis}
\label{cggta}

At room temperature the orthorhombic crystal structure with lattice parameters 
{\em a}=4.81 \AA, {\em b}=8.47 \AA,    and {\em c}=2.941 \AA\      and space group {\em Pbmm} 
($x\|a$, $y\|b$, $z\|c$) or, 
equivalently,  {\em Pmma} ($x\|b$, $y\|c$, $z\|a$) in standard setting, was proposed for 
CuGeO$_3$.\cite{vollenkle} The building  blocks of the structure are 
edge-sharing CuO$_6$ octahedra and corner-sharing GeO$_4$ tetrahedra stacked along the 
{\em c} axis of the crystal and resulting in Cu$^{2+}$ and Ge$^{4+}$ chains parallel to the 
{\em c} axis. These 
chains are linked together via the O atoms, and form layers parallel to the {\em b-c} plane weakly 
coupled along the {\em a} axis (Fig.\ \ref{cgstru1}). The unit cell contains 2 formula units 
of  CuGeO$_3$ (Fig.\ \ref{cgstru2}), 
with site group $C_{2h}^{y}$ for the 2 Cu atoms, $C_{2v}^{z}$ for the 2 Ge 
and the 2 O(1) atoms, and $C_{s}^{xz}$ for the 4 O(2) atoms [where O(2) denotes the O 
atoms linking the chains together].\cite{hirota,braden} A group theoretical analysis can be 
performed, working in standard orientation, to obtain the number and the symmetry of the lattice 
vibrational modes. Following the nuclear site group analysis method extended to 
crystals,\cite{rousseau} the contribution of each occupied site to the total irreducible 
representation of the crystal is:
\begin{eqnarray}
 &&\Gamma_{\rm{Cu}}\!=\!A_{u}\!+\!2B_{1u}\!+\!B_{2u}\!+\!2B_{3u}\nonumber \, , \\  
 &&\Gamma_{\rm{Ge+O(1)}}\!=\!2[A_{g}\!+\!B_{1u}\!+\!B_{2g}\!
        +\!B_{2u}\!+\!B_{3g}\!+\!B_{3u}]\nonumber \,  , \\
 &&\Gamma_{\rm{O(2)}}\!=\!2A_{g}\!+\!A_{u}\!+\!B_{1g}\!+\!2B_{1u}\!+\!2B_{2g}\!+\!B_{2u}\!
        +\!B_{3g}\!+\!2B_{3u}\nonumber \, .  
\end{eqnarray}
Subtracting the silent modes (2$A_{u}$) and the acoustic modes 
($B_{1u}+B_{2u}+B_{3u}$), the irreducible representation of the optical vibrations 
in standard setting ({\em Pmma}), is:
\begin{figure}[t]
\centerline{\epsfig{figure=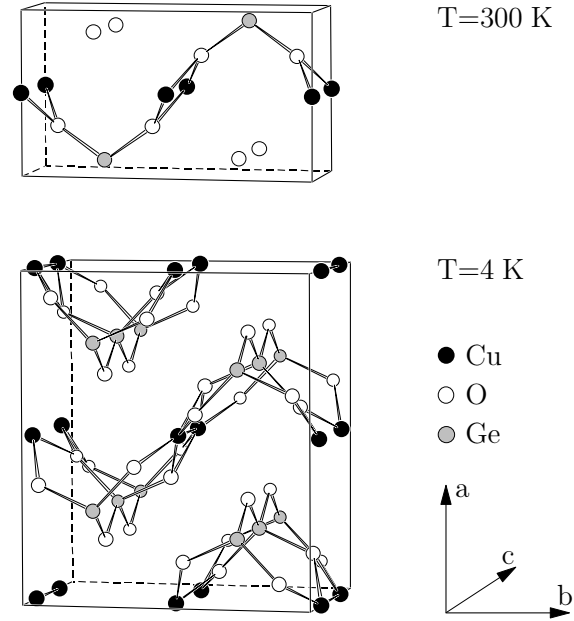,width=7.5cm,clip=}}
\vspace{.3cm}
\caption{Conventional unit cell of  CuGeO$_3$ in the undistorted 
(top) and SP phase (bottom). For clarity purpose, the ion displacements
due to the SP transition have been enlarged by a factor of 30.}
\label{cgstru2}
\end{figure}
\begin{eqnarray}
 \hspace{-.5cm}\Gamma\!=\! &&\,4A_{g}(aa,bb,cc)\!+\!B_{1g}(bc)\!+\!4B_{2g}(ab)\!
        +\!3B_{3g}(ac)\nonumber \\
 &&+5B_{1u}(E\|a)\!+\!3B_{2u}(E\|c)\!+\!5B_{3u}(E\|b) \, .
\end{eqnarray}
This corresponds to an expectation of 12 Raman active modes 
($4A_{g}+B_{1g}+4B_{2g}+3B_{3g}$) and 13 infrared active modes 
($5B_{1u}+3B_{2u}+5B_{3u}$) for CuGeO$_3$, 
in agreement with the calculation done by Popovi\'c {\em et al.}\cite{popovic} 

At temperatures lower than $T_{\rm{SP}}$ the proposed crystal structure is still orthorhombic, 
but with lattice 
parameters $a^\prime\!=\!2\times a$, $b^\prime\!=\!b$ and $c^\prime\!=\!2\times c$ and space group 
{\em Bbcm} ($x\|a$, $y\|b$, $z\|c$) 
or , equivalently, {\em Cmca} ($x\|c$, $y\|a$, $z\|b$)  in standard setting.\cite{hirota,braden} 
The distortion of the lattice 
taking place at the phase transition (Fig.\ \ref{cgstru2}) is characterized by the dimerization of  Cu-Cu pairs along the 
{\em c} axis (dimerization out of phase in neighboring chains), together with a rotation of 
the GeO$_4$ 
tetrahedra around the axis defined by the O(1) sites (rotation opposite in sense 
for neighboring tetrahedra). Moreover, the O(2) sites of the undistorted structure split in an 
equal 
number of O(2a) and O(2b) sites, distinguished by the distances O(2a)-O(2a) and O(2b)-O(2b) shorter and 
larger than O(2)-O(2),\cite{braden} respectively. The SP transition is also characterized 
(Fig.\ \ref{cgstru2}) by 
a doubling of the unit cell (corresponding to a doubling of the degrees of freedom from 30 to 60). 
The site groups in the new unit cell are: $C_{2}^{x}$ for Cu, $C_{2}^{y}$ for O(1), and 
$C_{s}^{yz}$ for Ge, O(2a) and O(2b).\cite{braden}  Repeating the group theoretical 
analysis, we obtain for the contributions to the total irreducible representation:
\begin{eqnarray}
  &&\Gamma_{\rm{Cu}}\!=\!A_{g}\!+\!A_{u}\!+\!2B_{1g}\!+\!2B_{1u}\!+\!2B_{2g}\!+\!2B_{2u}\!
        +\!B_{3g}\!+\!B_{3u}\nonumber\, , \\
  &&\Gamma_{\rm{O(1)}}\!=\!A_{g}\!+\!A_{u}\!+\!2B_{1g}\!+\!2B_{1u}\!+\!B_{2g}\!+\!B_{2u}\!
        +\!2B_{3g}\!+\!2B_{3u}\nonumber\, ,\\
  &&\Gamma_{\rm{Ge+O(2a)+O(2b)}}\!=\!3[2A_{g}\!+\!A_{u}\!+\!B_{1g}\!+\!2B_{1u}\!+\!B_{2g}\nonumber\\
        &&\hspace{4.47cm}+\!2B_{2u}\!+\!2B_{3g}\!+\!B_{3u}]\nonumber\, .  
\end{eqnarray}
The irreducible representation of the optical vibrations of  CuGeO$_3$ in the SP phase in standard 
setting ({\em Cmca}), is:
\begin{eqnarray}
 \hspace{-.5cm}\Gamma_{\rm{SP}}\!=\!&&8A_{g}(aa,bb,cc)\!+\!7B_{1g}(ac)\!+\!6B_{2g}(bc)\!
                                                  +\!9B_{3g}(ab)\nonumber\\
                &&+9B_{1u}(E\|b)\!+\!8B_{2u}(E\|a)\!+\!5B_{3u}(E\|c) \, .
\end{eqnarray}
Therefore 30 Raman active modes ($8A_{g}+7B_{1g}+6B_{2g}+9B_{3g}$) 
and 22 infrared active modes ($9B_{1u}+8B_{2u}+5B_{3u}$) 
are expected  for  CuGeO$_3$ in the SP phase, all the additional vibrations 
being zone boundary modes activated by the folding of the Brillouin zone.

In order to compare the results obtained for the undistorted and the SP phase of CuGeO$_3$, it is better 
to rewrite the irreducible representations $\Gamma$ and $\Gamma_{\rm{SP}}$ into {\em Pbmm} and 
{\em Bbcm} settings, respectively, because both groups are characterized by: $x\|a$, $y\|b$ and 
$z\|c$. This can be done by permuting the  ($1g,2g,3g$) and ($1u,2u,3u$) indices  in such a way 
that  it corresponds  to the permutations of the axis relating {\em Pmma} to {\em Pbmm}, and 
{\em Cmca} to {\em Bbcm}. Therefore, the irreducible representations of the optical vibrations of 
CuGeO$_3$, for $T\!>\!T_{\rm{SP}}$ ({\em Pbmm}) and $T\!<\!T_{\rm{SP}}$ ({\em Bbcm}), respectively, are:
\begin{eqnarray}
   \hspace{-2cm}\Gamma^{\prime}\!=\!&&4A_{g}(aa,bb,cc)\!+\!4B_{1g}(ab)\!+\!3B_{2g}(ac)\!
                                +\!B_{3g}(bc)\nonumber\\
                        &&+3B_{1u}(E\|c)\!+\!5B_{2u}(E\|b)\!+\!5B_{3u}(E\|a)\, ,
\label{cgrepU}
\end{eqnarray}
\begin{eqnarray}
  \hspace{-.5cm}\Gamma^{\prime}_{\rm{SP}}\!=\!&&8A_{g}(aa,bb,cc)\!+\!9B_{1g}(ab)\!+\!7B_{2g}(ac)\!
                                +\!6B_{3g}(bc)\nonumber\\
                        &&+5B_{1u}(E\|c)\!+\!9B_{2u}(E\|b)\!+\!8B_{3u}(E\|a) \, .
\label{cgrepD}
\end{eqnarray}
It is now evident that the number of infrared active phonons  is expected to increase from 5 to 8,  
5 to 9 and 3 to 5 for  light polarized along the {\em a}, {\em b} and {\em c} axes, respectively. 

\section{Experimental}
\label{cgexp}

We investigated the optical properties of several 
Cu$_{1-\delta}$Mg$_{\delta}$GeO$_3$  (with $\delta$\,=\,0,\,0.01), and  
CuGe$_{1-x}${\em B}$_{x}$O$_3$ [with {\em B}=Si ({\em x}\,=\,0,\,0.007,\,0.05,\,0.1), and Al 
({\em x}\,=\,0,\,0.01)] single crystals, in the frequency range 20$\,$-$\,$32\,000 cm$^{-1}$. 
These high-quality single crystals, several centimeters long 
in the {\em a} direction, were grown from the melt by a floating zone technique.\cite{revcolevschi}  
Plate-like samples were easily cleaved perpendicularly to the {\em a} axis. Typical dimensions were 
about 2 and 6 mm parallel to the {\em b} and {\em c} axis, respectively. The thickness was chosen 
in dependence of the experiment to be performed. In reflectivity measurements, when enough material 
was available, several millimeters thick samples were used in order to avoid interference fringes 
in the spectra, due to Fabry-Perot resonances.\cite{klein} Particular attention had to be paid 
when measuring reflectivity in frequency regions characterized by weak excitations (i.e., 
from 1000 to 25\,000 cm$^{-1}$): Because of the multiple reflections within the sample 
these excitations, 
which would not be directly detectable in reflectivity, would be observable as an absorption with 
respect to the background dominated by the interference fringes. If a 
Kramers-Kronig transformation is performed on such a pathological data in order to obtain the 
optical conductivity, unphysical results would be produced. As a matter of fact, it is precisely 
for this reason that a charge-transfer excitation at 1.25 eV (10\,000 cm$^{-1}$) was erroneously 
reported in one of 
the early optical papers on CuGeO$_3$.\cite{terasaki} On the other hand, in transmission measurements, 
where interference fringes cannot be avoided, the  thickness of the sample was 
adjusted to the strength of the particular excitation under investigation.

The samples were aligned by conventional Laue diffraction and mounted in a liquid He flow 
cryostat to study the temperature dependence of the optical properties between 
4 and 300 K. Reflectivity and transmission measurements in the frequency range going from 
20\,-\,7000 cm$^{-1}$  were  performed with a Fourier 
transform spectrometer (Bruker IFS 113v), with polarized light, in order to probe the optical 
response of the crystals along the {\em b} and the {\em c} axes. In reflectivity a  near normal 
incidence configuration ($\theta\!=\!11^{\circ}$) was used. The absolute  reflectivity and 
transmission values  
were obtained by calibrating the data acquired on the samples against a gold mirror and an empty 
sample holder, respectively. For frequencies higher than 6000 cm$^{-1}$ a Woollam (VASE) 
ellipsometer was used in both transmission and reflection operational modes. Unfortunately, on 
this last system it was not possible to perform temperature dependent measurements.
The optical conductivity spectra were obtained by Kramers-Kronig transformations in the regions 
where only reflectivity spectra were measurable, and by direct inversion of the Fresnel 
equations wherever both reflection and transmission data were available.\cite{klein}
\begin{figure}[t]
\centerline{\epsfig{figure=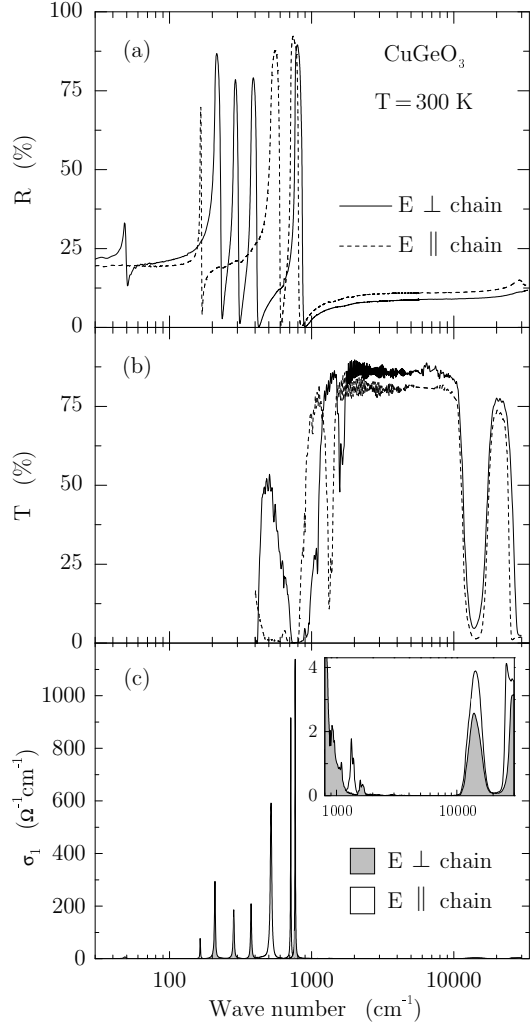,width=7cm,clip=}}
\vspace{.05cm}
\caption{Optical spectra of  CuGeO$_3$ at 300 K for  $E\!\parallel\!b$ 
(i.e., $\perp$ to the chain direction) and $E\!\parallel\!c$ (i.e., $\parallel$ to the chain 
direction), in the frequency range going from 30 to 34\,000 cm$^{-1}$.
 Panel (a), (b), and (c) show the results for reflectivity, transmission, and optical 
 conductivity, respectively. In the inset of panel (c) an enlarged  view of $\sigma_{1}(\omega)$ 
 from 800 to 30\,000 cm$^{-1}$ is presented (note the very low absolute value of 
 conductivity).}
\label{cgbc300}
\end{figure}

\section{Pure CuGeO$_3$}
\label{cgp}

Let us now, as an introduction to what we will discuss in detail in the following sections, 
describe briefly the main features of the optical spectra of pure  CuGeO$_3$ over the entire 
frequency range we covered  with our experimental systems.
In Fig.\ \ref{cgbc300} we present reflectivity, transmission, and conductivity spectra 
of  CuGeO$_3$ at 300 K, for  $E\!\parallel\!b$ \,(i.e., $\perp$ to the chain direction), and 
$E\!\parallel\!c$  \,(i.e., $\parallel$ to the chain direction), in the frequency range 
going from 30 to 34\,000 cm$^{-1}$. The results are typical of an ionic insulator.
The far-infrared region ($\omega\!<$1000 cm$^{-1}$) is characterized by strong optical phonon 
modes showing the expected anisotropy for the {\em b} and {\em c} axes, as one can see from 
 reflectivity and conductivity  (the detail discussion 
of the phonon spectra will be subject of the next section). Besides the phonon lines, no 
background conductivity is observable (Fig.\ \ref{cgbc300}c). Transmission spectra for 
$\omega\!<$400 cm$^{-1}$ are not shown. 
However, infrared transmission measurements carried out at low temperature, in order 
to investigate very weak magnetic and lattice excitations, will be discussed later in the 
course of the paper. At frequencies larger than 1000 cm$^{-1}$, reflectivity is low and 
almost completely featureless. More information can be gained from transmission in this case.
In Fig.\ \ref{cgbc300}b we can see (in addition to the strong phonons for $\omega\!<\!1000$ 
cm$^{-1}$) absorption processes at 1330 and 1580 cm$^{-1}$ along the  {\em c} and {\em b} axes, 
respectively, and at $\sim$14\,000 and $\sim$27\,000 cm$^{-1}$, with approximately the same 
frequency for the two different axes. Having both reflectivity and transmission data in this 
region, we could calculate the dynamical conductivity by direct inversion of the Fresnel 
formula.\cite{klein} As shown by Fig.\ \ref{cgbc300}c and, in particular, 
by the enlarged view given in the inset (note the very low absolute value of conductivity), 
extremely weak excitations are present, in the region going from 1000 to 30\,000 
cm$^{-1}$, on top of a zero background. Let us now briefly discuss the nature of these 
excitations. At 1000 cm$^{-1}$ we can see the vanishing tail of the highest {\em b}-axis phonon 
(inset of Fig.\ \ref{cgbc300}c). Above that, we find the two peaks at 1330 and 1580 cm$^{-1}$ 
along the  {\em c} and {\em b} axes, respectively. On the basis of the energy position and of the 
temperature dependence, these features can be ascribed to multiphonon processes. At 
$\sim$14\,000 cm$^{-1}$ ($\sim$1.8 eV), for both orientations of the electric field, a very 
weak peak is present which has been shown to be due to  phonon-assisted Cu {\em d-d} 
transitions.\cite{bassi} Finally, the onset of the Cu-O charge-transfer excitations is observable 
at $\sim$27\,000 cm$^{-1}$. Superimposed to it are some sharper features of probable excitonic 
nature.
\begin{figure}[t]
\centerline{\epsfig{figure=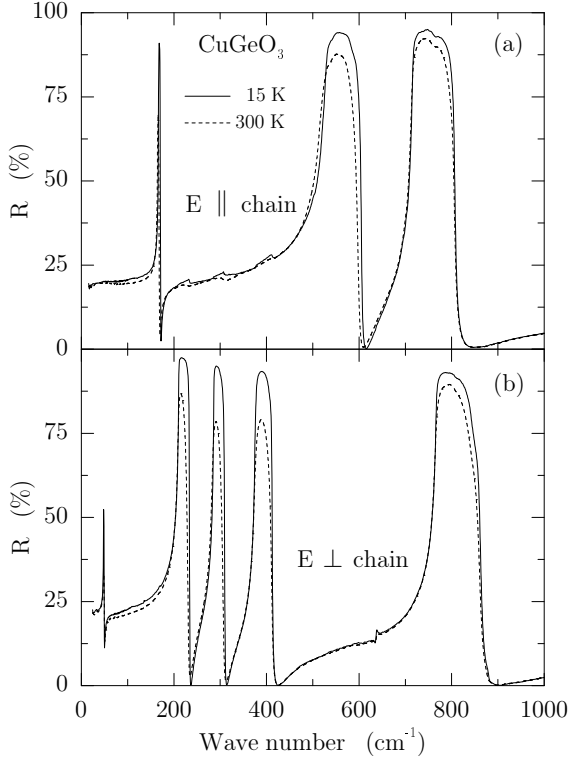,width=7.5cm,clip=}}
\vspace{.3cm}
\caption{Reflectivity of a single crystal of pure CuGeO$_3$, as a function of frequency,  
at two different temperatures (300 K and 15 K) in the undistorted phase. The spectra 
are shown for $E$ parallel and perpendicular to the chain direction in panel (a) and (b), 
respectively.}
\label{cgbcref1}
\end{figure}

\subsection{Phonon Spectrum and Lattice Distortion}
\label{cgphdist}

In this section we will discuss in detail the phonon spectrum of pure CuGeO$_3$, for both the high 
temperature undistorted phase and the low temperature dimerized SP phase. In particular, we 
will analyze the optical data in order to find possible zone-boundary folded modes activated by 
the lattice distortion involved in the SP phase transition. The {\em c} and {\em b}-axis 
({\em i.e.}, $E\!\parallel$\,chain and $E\!\perp$\,chain, respectively) reflectivity 
spectra of CuGeO$_3$ in the undistorted phase are presented in Fig.\ \ref{cgbcref1}, for 
two different temperatures higher than $T_{\rm{SP}}$=14 K. 
\begin{figure}[t]
\centerline{\epsfig{figure=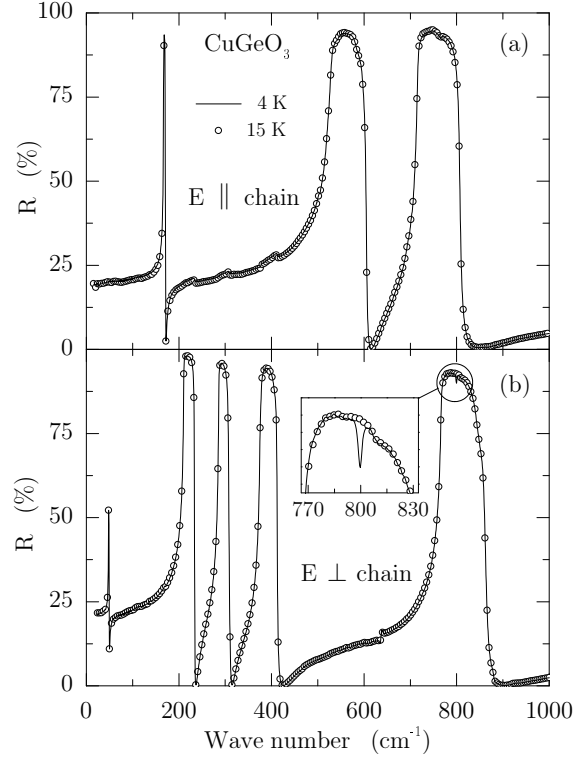,width=7.5cm,clip=}}
\vspace{.3cm}
\caption{Comparison between reflectivity spectra measured in the SP phase at 4 K 
(solid line) and just before the SP transition at 15 K (circles) on a pure single-crystal of
CuGeO$_3$. For $E\!\parallel$\,chain (a) no difference is found across 
the phase transition whereas for  $E\!\perp$\,chain (b) a new feature appears 
at 800 cm$^{-1}$ (as clearly shown in the inset).}
\label{cgbcref2}
\end{figure}
The data are shown up to 1000 cm$^{-1}$ which covers  the full 
phonon spectrum. Three phonons are detected  along the {\em c} axis 
($\omega_{\rm{TO}}\approx$\,167, 528 and 715 cm$^{-1}$, for $T\!=\!15$ K), and five along the 
{\em b} axis ($\omega_{\rm{TO}}\approx$\,48, 210, 286, 376 and 766  cm$^{-1}$, 
for $T\!=\!15$ K). This is in agreement with what is expected on the basis of the 
group-theoretical analysis presented in section\ \ref{cggta}, (see Eq.\ \ref{cgrepU}) for the  
space group {\em Pbmm} proposed for  CuGeO$_3$ in the uniform phase.\cite{vollenkle} The 
structure in 
Fig.\ \ref{cgbcref1}a between 200 and 400 cm$^{-1}$  is due to a leakage of the polarizer 
and corresponds to the three modes detected along the {\em b} axis in the same frequency range. 
Similarly, the feature at approximately 630 cm$^{-1}$  in Fig.\ \ref{cgbcref1}b is a leakage of 
a mode polarized along the {\em a} axis.\cite{popovic} However, the reason why this phonon has 
been detected in the {\em b}-axis reflectivity is not simply, as in the previous case, a leakage 
of the polarizer. It 
has to do with the finite angle of incidence $\theta\!=\!11^{\circ}$ of the radiation on the 
sample, and with the fact that {\em p}-polarized light was used to probe the optical response along 
the {\em b} axis. In fact, whereas for {\em s}-polarized light the electric field was parallel 
to the {\em b-c} plane, in {\em p} polarization there was a small but finite component of the 
electric vector perpendicular to the plane of the sample, which could then couple to the 
{\em a}-axis excitations (at least to those particularly intense). As a last remark, we would like 
to stress  that the temperature dependence observable in the reflectivity spectra of 
Fig.\ \ref{cgbcref1} mainly corresponds to the hardening and sharpening 
of the phonon lines, usually observable in a crystalline material upon reducing the temperature. 
However, a more extensive discussion of the temperature dependence of the phonon parameters 
over a broad temperature range will be presented in section\ \ref{cgppara}, in relation to 
the soft-mode issue.

\begin{figure}[t]
\centerline{\epsfig{figure=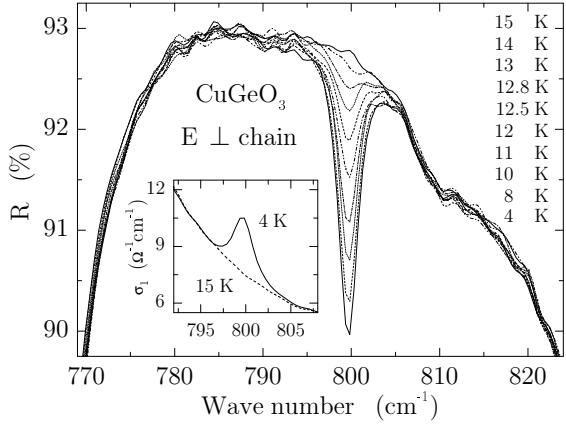,width=7.5cm,clip=}}
\vspace{.3cm}
\caption{Detailed temperature dependence of the 800 cm$^{-1}$ line observed with 
$E\!\perp$\,chain in the reflectivity  spectra for $T\!<\!T_{\rm{SP}}$. 
In the inset, where the dynamical conductivity calculated via Kramers-Kronig analysis is plotted, 
a new peak is clearly visible for $T\!=\!4$ K.}
\label{cgsp1}
\end{figure}

Cooling down the sample below the phase transition temperature $T_{\rm{SP}}$=14 K, we 
can seek for changes in the phonon spectrum with respect to the result obtained at temperature 
just above $T_{\rm{SP}}$. The {\em c} and {\em b}-axis reflectivity spectra 
measured at $T\!=\!15$ K (circles) and $T\!=\!4$ K (solid line) are compared in Fig.\ \ref{cgbcref2}.  
The 15 K curves have been plotted with lower resolution, for the sake of clarity. 
The solid line is the 4 K experimental result.
Whereas for $E\!\parallel\!c$ the spectra are exactly identical, a new feature, though very weak, 
is detected in the SP phase  at 800 cm$^{-1}$  for $E\!\parallel\!b$, as shown in the inset 
of  Fig.\ \ref{cgbcref2}b.  A careful investigation of temperatures ranging from 4 to 15 K 
(see Fig.\ \ref{cgsp1}) clearly shows that this feature, that falls in the frequency region 
of high reflectivity for the $B_{2u}$-symmetry mode at 766 cm$^{-1}$ and
therefore shows up mainly for its absorption, is activated by the SP 
transition.\cite{sp} It  corresponds to a new peak in conductivity, superimposed on the  
background due to the lorentzian tail of the close $B_{2u}$ phonon 
(see inset of Fig.\ \ref{cgsp1}). 

\begin{figure}[t]
\centerline{\epsfig{figure=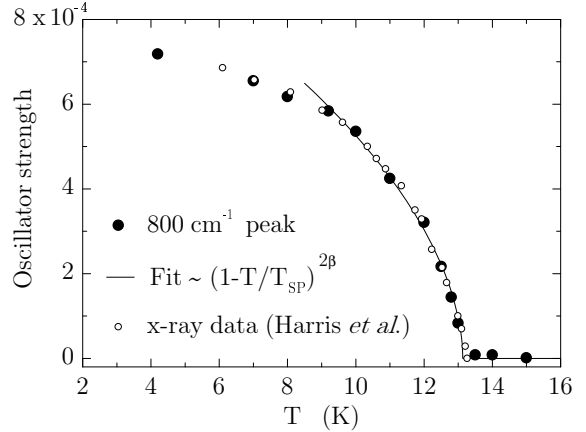,width=7.5cm,clip=}}
\vspace{.3cm}
\caption{Oscillator strength of the 800 cm$^{-1}$ folded mode, 
observed for $E\!\perp$\,chain on a pure CuGeO$_3$ single crystal, plotted versus temperature. 
The x-ray scattering data of Harris et al.\protect\cite{harris}  
and the fit to a power law  in the reduced temperature are also shown.}
\label{cgsp2}
\end{figure}

By fitting the reflectivity spectra with Lorentz oscillators for all the optical phonons, it is 
possible to obtain the temperature dependence  of the oscillator strength for the 800 cm$^{-1}$ 
feature. The results are plotted in Fig.\ \ref{cgsp2}, together with the peak intensity of a 
superlattice reflection measured by Harris {\em et al.}\cite{harris} in an x-ray scattering 
experiment on a pure CuGeO$_3$  single crystal characterized, as our sample, by 
$T_{\rm{SP}}$$\approx$13.2 K. The data clearly show the second-order character of 
the phase 
transition. From the perfect agreement of the infrared  and x-ray scattering 
results, and from the observation that the resonant frequency is not shifting at all with 
temperature, we can conclude that the peak at 800 cm$^{-1}$ corresponds to a pure lattice 
excitation. It has to be a folded zone-boundary mode most probably related to the $B_{2u}$  phonon 
observed at  
766 cm$^{-1}$, which is mainly an oxygen vibration.\cite{popovic}  We can then estimate for the 
$B_{2u}$ mode  an energy dispersion, over the full Brillouin zone, of the order of 
34 cm$^{-1}$=4.22 meV, at $T\!=\!15$ K. 
If magnetic degrees of freedom were involved 
(e.g., if this excitation were a magnon-plus-phonon process\cite{lorenzana1}), not only 
a decrease in intensity would be observable, but also a frequency shift 
to lower values, reflecting the closing of the magnetic gap 
for $T\!\rightarrow\!T_{\rm{SP}}$ from below. In fact, on the basis of Cross and Fisher 
theory,\cite{crossfi} the magnetic gap would scale as $\delta^{2/3}$  
[where $\delta\!\sim\!(1-T/T_{\rm{SP}})^{\beta}$, close to $T_{\rm{SP}}$, denotes the 
generalize symmetry-breaking lattice distortion]. Its gradual closing has been 
directly observed in neutron and inelastic light scattering 
experiments.\cite{harris1,els} 

Because both the intensity of the superlattice reflections measured with x-ray or neutron 
scattering, and the intensity of the zone-boundary folded modes in an optical experiment are  
proportional to $\delta^2$ (see Ref.\ \onlinecite{bruce}), we can try to fit the temperature 
dependence of the 
oscillator strength for the 800 cm$^{-1}$ SP-activated mode to the equation 
$\sim(1-T/T_{\rm{SP}})^{2\beta}$. As a result of the fit performed over a broad temperature 
range (see Fig.\ \ref{cgsp2}), we obtained $\beta\!=\!0.26\pm0.02$,  in agreement with 
Ref.\ \onlinecite{harris1}. However, the best fit  value of  $\beta$  is strongly dependent 
on the temperature range chosen to fit the data.\cite{lumsden} If only 
points very close (within 1 K)  to $T_{\rm{SP}}$  are considered,  the value
$\beta\!=\!0.36\pm0.03$ is obtained, as reported in Ref.\ \onlinecite{lumsden1}.  At this point, 
in relation to the soft-mode issue in  CuGeO$_3$ treated in section\ \ref{cgppara}, it is  worth 
mentioning that for the organic SP system TTF-CuS$_4$C$_4$(CF$_3$)$_4$, which shows a precursive 3D 
soft-phonon at the superlattice position, the value $\beta\!=\!0.5$ was obtained.\cite{moncton} 
It was argued by Cross and Fisher,\cite{crossfi} that this soft-mode 
is responsible for the mean-field behavior, i.e. $\beta\!=\!1/2$, in the TTF salt.

\begin{figure}[t]
\centerline{\epsfig{figure=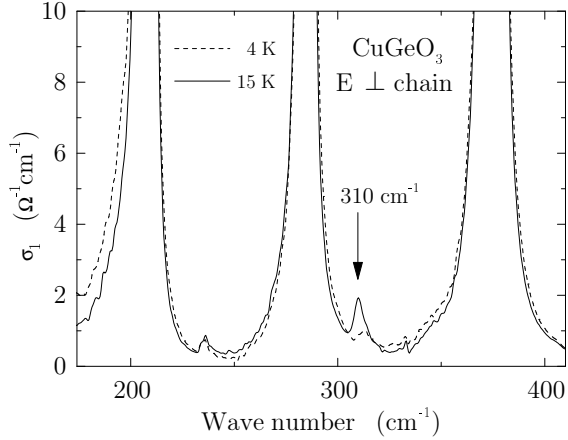,width=7.5cm,clip=}}
\vspace{.3cm}
\caption{Comparison between conductivity spectra measured in the SP phase at 4 K 
and just before the SP transition at 15 K, for $E\!\perp$\,chain, on a pure single-crystal of
CuGeO$_3$: A zone boundary folded mode  appears, across 
the phase transition, at 310 cm$^{-1}$.}
\label{cgspcon}                          
\end{figure}
\begin{figure}[t]
\centerline{\epsfig{figure=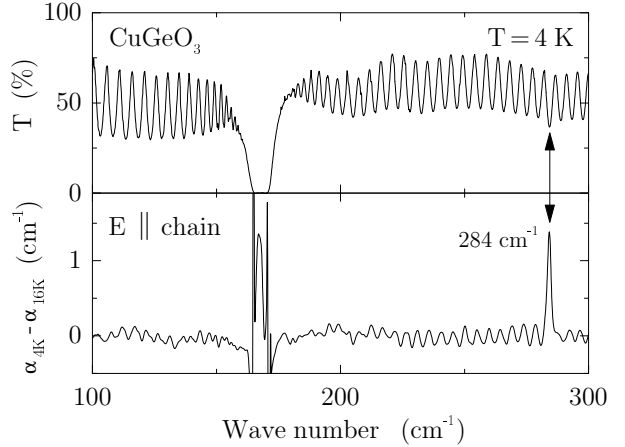,width=8cm,clip=}}
\vspace{.3cm}
\caption{Top panel: Transmission spectrum of a pure single-crystal of  CuGeO$_3$  measured in 
the SP phase, at 4 K, with $E\!\parallel$\,chain. Bottom panel: The absorbance difference 
spectrum  $\alpha_{4K}-\alpha_{16K}$ clearly shows an activated zone boundary mode at 284 
cm$^{-1}$.}
\label{cgsptra}
\end{figure}

As we saw in  Fig.\ \ref{cgbcref2}, only one phonon activated by the SP phase transition, of the 
many we actually expected, is directly observable in reflectivity spectra. However, more lines 
can be found by means of a deeper analysis, 
in agreement with the results reported in Ref.\ \onlinecite{popova}. In fact, by checking the 
optical conductivity at 4 and 15 K, a second folded mode is found along the 
{\em b} axis at 310 cm$^{-1}$ (see Fig.\ \ref{cgspcon}). This line was not distinguishable 
in reflectivity because it coincides with $\omega_{\rm{LO}}$ of the $B_{2u}$-symmetry phonon at 
286 cm$^{-1}$. The reasons for not detecting all the phonons predicted from the group 
theoretical analysis (Eq.\ \ref{cgrepD}) in reflectivity measurements, at $T\!<\!T_{\rm{SP}}$,  
are probably the small values 
of the atomic displacements involved  in the SP transition (with a correspondingly small 
oscillator strength of zone-boundary modes), and/or  possibly the small dispersion of the 
optical branches of some of the lattice vibrations. However, it is  not surprising that the 
only activated modes have been detected along the {\em b} axis of the crystal: for this particular 
direction, below the phase transition, a strong spontaneous thermal contraction has been 
observed,\cite{harris1} which can be responsible for a relative increase of the oscillator 
strength of the phonons polarized along the  {\em b} axis, with respect to those polarized 
along the {\em c} axis. 

In order to probe the phonon spectrum of the dimerized phase with a more sensitive tool, 
we performed transmission measurements on a 350 $\mu$m thick sample, in the far-infrared 
region. One more line was then detected at 284 cm$^{-1}$ along the {\em c} axis, as shown in 
Fig.\ \ref{cgsptra}.  In the top panel a transmission spectrum acquired 
at 4 K is plotted in the frequency range 100-300 cm$^{-1}$. Fabry-Perot interference fringes 
are clearly visible, interrupted at $\sim$167 cm$^{-1}$ by the strong absorption of the lowest 
energy $B_{1u}$-symmetry phonon. Moreover,  at 284 cm$^{-1}$ one can observe a slightly 
more pronounced 
minimum in the interference pattern. However, because of the weakness of this line, a real peak 
can be observed only in the absorbance difference spectrum (bottom panel of Fig.\ \ref{cgsptra}). 
The spikes present at $\sim$167 cm$^{-1}$ in the absorbance difference spectrum
are due to the complete absorption of the light in that frequency range (top panel). As a 
last remark, we want to stress that the final assignment of the 310 and 284 cm$^{-1}$ peaks 
to optical phonons activated by the SP transition is based, as for the line at 
800 cm$^{-1}$, on their  detailed temperature dependence (not shown). 

\subsection{Phonon Parameters and Soft-Mode Issue}
\label{cgppara}

\begin{figure}[t]
\centerline{\epsfig{figure=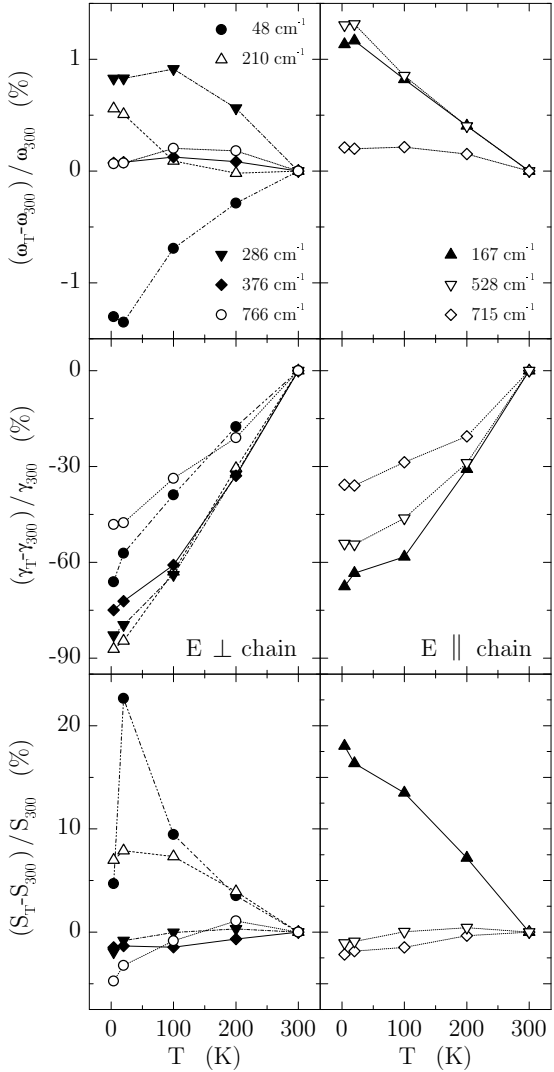,width=7.3cm,clip=}}
\vspace{.15cm}
\caption{Temperature dependence of the percentage change in resonant frequency, damping, 
and oscillator strength (from top to bottom) of the optical phonons of the 
undistorted phase, observed for $E\!\perp$\,chain (left) and $E\!\parallel$\,chain (right) on the 
pure single-crystal of CuGeO$_3$.}
\label{cgpara}
\end{figure}

As far as the dynamical interplay between spins and phonons in CuGeO$_3$  is concerned, it is 
clear from the reflectivity spectra plotted in Fig.\ \ref{cgbcref1} and\ \ref{cgbcref2} 
that a well-defined soft mode, driving the structural deformation in CuGeO$_3$, has not been 
detected in our measurements. However, as any dimerization must be related to normal modes away 
from the zone center, the softening of one or more modes, across the SP phase transition, 
should be expected at $k\!=\!(\pi\!/\!a$,0,$\pi\!/\!c)$, the actual propagation vector in CuGeO$_3$. 
As optical techniques can probe the phonon branches only at the $\Gamma$ 
point ($k\!=\!0$), the softening can, strictly speaking, be investigated only by neutron 
scattering. Nevertheless, we tried to gain interesting 
insights from the temperature dependence of the phonon parameters obtained from the fit
of the reflectivity data (see Fig.\ \ref{cgpara} and table\ \ref{cgparatab}). In fact, if the 
dispersion and the mixing 
of the phonon branches are not too strong, one could hope that the presence of a soft mode at 
$k\!=\!(\pi\!/\!a$,0,$\pi\!/\!c)$ would result in an overall softening of the branch it belongs 
to. Therefore,  our first attempt was to check if any of the optical phonons detected in 
our experiments was showing a red shift upon cooling the sample.
In Fig.\ \ref{cgpara} we can clearly observe that the $B_{2u}$  mode ($E\!\perp$\,chain) at 
48 cm$^{-1}$ is the only one showing an evident monotonic red shift from 300 to 15 K . 
Below $T_{\rm{SP}}$ this mode shows only a small blue shift. Also the behavior of the oscillator 
strength of this phonon is rather interesting: It grows continuously from 300 to 15 K (a $\sim$23\% 
increase), and it shows a sudden drop of $\sim$15\% across the phase transition.
On the basis of these results, we could conclude that these particular phonon branch could have 
been a good candidate to show a real softening at the SP distortion vector in the Brillouin zone.

\begin{table}[t]
\begin{tabular}{lcccccc} 
Mode     & Param.         &  4 K   &  15 K  & 100 K  & 200 K  & 300 K  \\ \hline  
         & $\omega_{\rm{TO}}$ & 166.78 & 166.82 & 166.25 & 165.56 & 164.90 \\
$B_{1u}$ & $\gamma$           & 0.41   & 0.46   & 0.53   & 0.88   & 1.27   \\
         & {\em S}            & 0.348  & 0.343  & 0.335  & 0.316  & 0.295  \\ 
         & $\omega_{\rm{TO}}$ & 527.69 & 527.74 & 525.33 & 522.98 & 520.89 \\
$B_{1u}$ & $\gamma$           & 4.07   & 4.05   & 4.78   & 6.32   & 8.89   \\
         & {\em S}            & 1.900  & 1.903  & 1.921  & 1.929  & 1.921  \\
         & $\omega_{\rm{TO}}$ & 715.25 & 715.17 & 715.27 & 714.84 & 713.75 \\
$B_{1u}$ & $\gamma$           & 3.12   & 3.11   & 3.46   & 3.85   & 4.85   \\
         & {\em S}            & 0.636  & 0.638  & 0.640  & 0.648  & 0.650  \\ \hline 
         & $\omega_{\rm{TO}}$ & 48.28  & 48.26  & 48.58  & 48.78  & 48.92  \\
$B_{2u}$ & $\gamma$           & 0.63   & 0.79   & 1.13   & 1.52   & 1.85   \\
         & {\em S}            & 0.286  & 0.335  & 0.299  & 0.282  & 0.273  \\
         & $\omega_{\rm{TO}}$ & 210.27 & 210.17 & 209.29 & 209.07 & 209.10 \\
$B_{2u}$ & $\gamma$           & 0.44   & 0.53   & 1.28   & 2.41   & 3.46   \\
         & {\em S}            & 1.717  & 1.732  & 1.723  & 1.669  & 1.605  \\
         & $\omega_{\rm{TO}}$ & 285.79 & 285.79 & 286.03 & 285.04 & 283.45 \\
$B_{2u}$ & $\gamma$           & 0.83   & 0.99   & 1.75   & 3.27   & 4.84   \\
         & {\em S}            & 0.765  & 0.773  & 0.780  & 0.782  & 0.780  \\
         & $\omega_{\rm{TO}}$ & 375.62 & 375.65 & 375.81 & 375.66 & 375.35 \\
$B_{2u}$ & $\gamma$           & 1.44   & 1.60   & 2.26   & 3.86   & 5.76   \\
         & {\em S}            & 0.596  & 0.597  & 0.596  & 0.601  & 0.605  \\
         & $\omega_{\rm{TO}}$ & 766.28 & 766.34 & 767.35 & 767.16 & 765.79 \\
$B_{2u}$ & $\gamma$           & 3.27   & 3.30   & 4.18   & 4.98   & 6.30   \\
         & {\em S}            & 0.677  & 0.689  & 0.705  & 0.718  & 0.711  \\ \hline\hline 
         & $\omega_{\rm{TO}}$ & 284.21  & - & - & - & - \\       
$B_{1u,{\rm{SP}}}$  & $\gamma$      & --- & - & - & - & - \\    
         & {\em S}            & --- & - & - & - & -  \\   \hline        
         & $\omega_{\rm{TO}}$ & 309.58  & - & - & - & - \\    
$B_{2u,{\rm{SP}}}$  & $\gamma$      & 4.27    & - & - & - & - \\    
         & {\em S}            & 0.003 & - & - & - & - \\
         & $\omega_{\rm{TO}}$ & 799.75  & - & - & - & - \\       
$B_{2u,{\rm{SP}}}$  & $\gamma$      & 2.32    & - & - & - & - \\    
         & {\em S}            & 0.0007 & - & - & - & - \\
\end{tabular}
\vspace{.15cm}
\caption{Resonant frequency $\omega_{\rm{TO}}$ (cm$^{-1}$), damping $\gamma$ (cm$^{-1}$), 
and oscillator strength {\em S} of the optical phonons detected for $E\!\parallel$\,chain ($B_{1u}$ 
symmetry) and $E\!\perp$\,chain ($B_{2u}$ symmetry) on the pure single-crystal of CuGeO$_3$. The 
parameters have been obtained by fitting the phonon spectra with Lorentz oscillators, at 
different temperatures.}
\label{cgparatab}
\end{table}  

In order to check more directly the possible presence of a soft mode in CuGeO$_3$, we performed 
transmission measurements with a mm-wave transmission setup, equipped with a backward wave 
oscillator operating in the frequency range 3.6\,-\,6.0 cm$^{-1}$.\cite{kozlov} The measurements 
were done on a $\sim$1 mm thick sample of pure  CuGeO$_3$  and, in order to relax the 
{\em k}-conservation rule and to obtain a result averaged over the all Brillouin zone, 
on Si and Mg substituted crystals, with approximately the same thickness of the pure one. 
In this way, we aimed to measure very accurately the transmission through the samples at 
one very low fixed frequency (5 cm$^{-1}$), as a function of temperature. If a mode would get 
soft, it would possibly result in a temperature dependent transmission showing a `w-shape' 
centered at $T_{\rm{SP}}$: Two minima in the absolute transmission, one on each side of  
$T_{\rm{SP}}$. Unfortunately, we could not detect any change. 
Had we seen the expected `w-shape' behavior, this would have 
been a strong indication of the existence of a soft mode in CuGeO$_3$. In the 
present case we cannot confirm nor rule out a soft mode.

The final word about this issue had to come, as expected, from neutron scattering measurements. 
And indeed, recently, it was clearly shown  that the pre-existing soft mode, expected in the 
classical theories of the SP phase transition,\cite{bulaevskii,crossfi} is not there in the 
case of CuGeO$_3$.\cite{braden1} Neither is present the central peak usually observed in 
order-disorder phase transitions.\cite{bruce} In fact, Braden {\em et al.} investigated the lattice 
dynamics of  CuGeO$_3$ with inelastic neutron scattering combined with shell model lattice dynamical 
calculations.\cite{braden1} They could identify the low-lying modes of the symmetry of the 
structural distortion characterizing the dimerized SP phase, and show that there is no soft 
mode behavior for these phonons at $k\!=\!(\pi\!/\!a$,0,$\pi\!/\!c)$. 
Moreover, they found that the static SP distortion 
does not correspond to one single normal mode of the uniform phase, as usually expected 
for a continuous displacive transition,\cite{bruce} but it corresponds to the linear combination, 
in ratio of 3:2, of two out of four optical phonons allowed by symmetry. Of these two modes, 
one involves an appreciable Cu displacement (essential for the dimerization), the second is 
associated with the twisting of the CuO$_2$ ribbons (which is the second element of the 
distortion). Finally, they showed that none of the two  belongs to a branch having,  
at $k\!=\!0$, an optically allowed symmetry.\cite{braden1} 
 
As a consequence of these findings, it appears that the behavior of the phonon parameters for 
the $B_{2u}$ mode 
at  48 cm$^{-1}$ is simply `accidental'. One may speculate that the absence of a 
softening at $k\!=\!(\pi\!/\!a$,0,$\pi\!/\!c)$  implies that the phase transition is not driven 
by a softening of the phonon spectrum at $k\!=\!(\pi\!/\!a$,0,$\pi\!/\!c)$, but only by a 
change in electronic structure which, in turn, determines the dynamical charge of the ions 
and the interatomic force constants. In this scenario, the large 
change in oscillator strength of some of the vibrational modes observed in our optical data 
(see Fig.\ \ref{cgpara}), results from a change in ionicity, or, in other words, a transfer 
of spectral weight from the elastic degrees of freedom to electronic excitations. As far as 
the understanding of the SP phase transition in  CuGeO$_3$ is concerned, now that the absence of a 
soft phonon is a well established experimental fact,\cite{braden1} a further discussion will be 
presented in section\ \ref{cgdisc}.

A last remark has to be made regarding the temperature dependence of the $B_{2u}$ mode 
observed at 286 cm$^{-1}$. On the basis of optical  reflectivity measurements,\cite{li}  
softening of this phonon, upon going through the phase transition, was suggested. This is not 
confirmed by our results which show no considerable frequency shift for this mode upon 
reducing the temperature from  15 to 4 K (see Fig.\ \ref{cgpara}). On the other hand a 
reduction of both the scattering rate and the oscillator strength is observed, which can 
explain the double-peak structure in the reflectance ratio $R$(20 K)/$R$(5 K) reported in 
Ref.\ \onlinecite{li}.

\subsection{Magnetic Excitations}
\label{singtrip}

\begin{figure}[t]
\centerline{\epsfig{figure=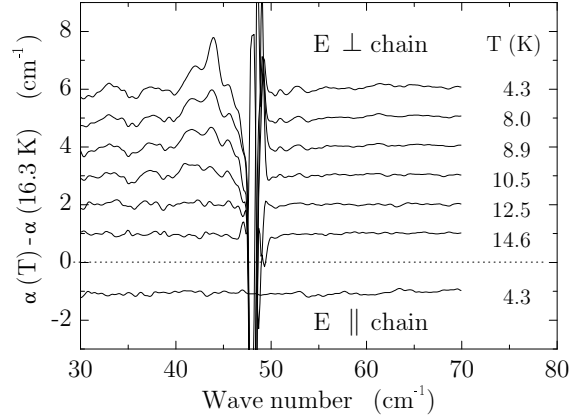,width=7.5cm,clip=}}
\vspace{.3cm}
\caption{Absorbance difference spectra  for the pure single-crystal of CuGeO$_3$, 
with  $E\!\perp$\,chain (top) and $E\!\parallel$\,chain (bottom). The spectra have been shifted 
for clarity.}
\label{cgtrip}
\end{figure}

Inelastic neutron and light scattering experiments are very sensitive to magnetic excitations and, 
in the case of CuGeO$_3$, they have been very powerful tools in detecting the singlet-triplet magnetic
gap,\cite{nishi,fujita} separated by a second gap from a continuum of magnon 
excitations,\cite{ain,arai,kuroe,loosdrecht1} and a singlet bound-states within the two energy 
gaps.\cite{els,loosdrecht2} On the other hand, optical spectroscopy is usually not the elective 
technique to study this kind of processes, unless a static (charged 
magnons\cite{anvprl,anvphysica}) or a dynamic (phonon assisted bi-magnons\cite{lorenzana1}) 
breaking of symmetry is present in the system under investigation. 
Moreover, in these two latter cases, only those excitations 
characterized by a total $\Delta S\!=\!0$ can be probed, because of spin conservation. Therefore, 
a direct singlet-triplet excitation is, in principle, 
not detectable. However, such a transition has been observed 
at 44.3 cm$^{-1}$ in an infrared transmission experiment where the singlet-triplet nature of the 
transition was demonstrated by the Zeeman splitting observed applying a magnetic 
field.\cite{loosdrecht3} 
In a later theoretical paper,\cite{uhrig0} this line was interpreted as a magnetic excitation 
across the gap at the wave vector (0,2$\pi\!/b$,0) in the Brillouin zone, activated by the 
existence of staggered magnetic fields along the direction perpendicular to the chains. 
Therefore, in a transmission experiment performed with polarized light incoming onto the 
{\em b-c} plane of the sample at normal incidence, this excitation would be 
detectable only for magnetic field of the radiation parallel to the {\em b} axis 
 of the crystal ($\perp$\,chain), i.e., $E$ parallel to the {\em c} axis 
($\parallel$\,chain).\cite{uhrig0} In order to verify this interpretation, 
we measured the far-infrared transmission on a 350 $\mu$m thick pure single crystal of CuGeO$_3$,  
for both $E\!\perp$\,chain and $E\!\parallel$\,chain. The absorbance difference spectra are 
reported in Fig.\ \ref{cgtrip}, where they have been shifted for clarity. 
Contrary to what expected following the interpretation given in Ref.\ \onlinecite{uhrig0}, 
for $E\!\parallel$\,chain (Fig.\ \ref{cgtrip}, bottom) no absorption is observed. However, for  
$E\!\perp$\,chain (Fig.\ \ref{cgtrip}, top) an absorption peak, showing the temperature 
dependence appropriate for an excitation activated by the SP phase transition, is present at 
approximately 44 cm$^{-1}$. One has to note that the feature at 48 cm$^{-1}$ is produced by 
the low-energy $B_{2u}$ phonon. In fact, because of the complete absorption of the light in 
that frequency range, spikes are obtained when absorbance difference spectra are calculated.
The observed  polarization dependence puts a strong experimental constraint on the possible 
microscopic mechanism giving rise to the singlet-triplet absorption peak. For a complete 
understanding, a full experimental investigation of the symmetry of this magnetic absorption 
process by ,e.g., measuring transmission through the {\em a-b} and {\em a-c} planes for  
different orientations of the electric and magnetic fields, is needed. A discussion of this 
feature, in relation to other critical theoretical and experimental aspects of the physics of 
CuGeO$_3$, will be presented in section\ \ref{cgdisc}.
\begin{figure}[t]
\centerline{\epsfig{figure=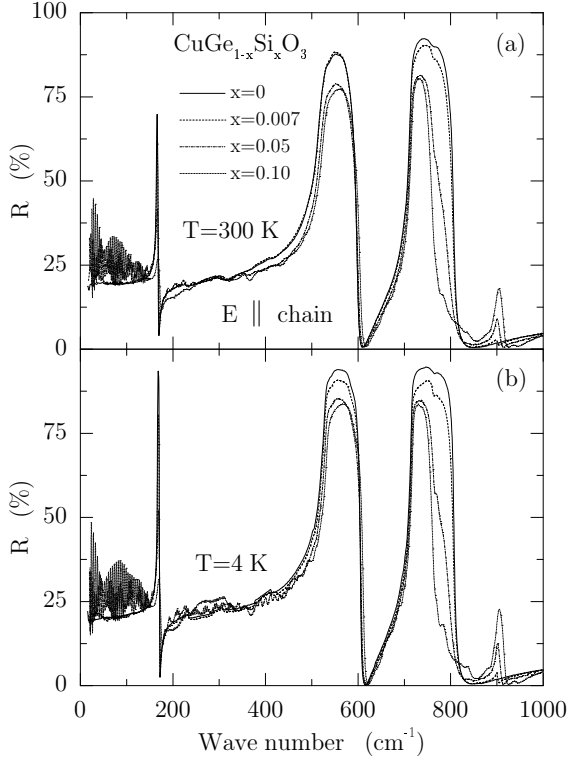,width=7.5cm,clip=}}
\vspace{.3cm}
\caption{$c$-axis reflectivity spectra of Si-doped single crystals of CuGeO$_3$,  
for different silicon concentrations, at two different temperatures: $T\!=\!300$ K (a), 
and $T\!=\!4$ K (b).}
\label{cgdref1}
\end{figure}
\begin{figure}[t]
\centerline{\epsfig{figure=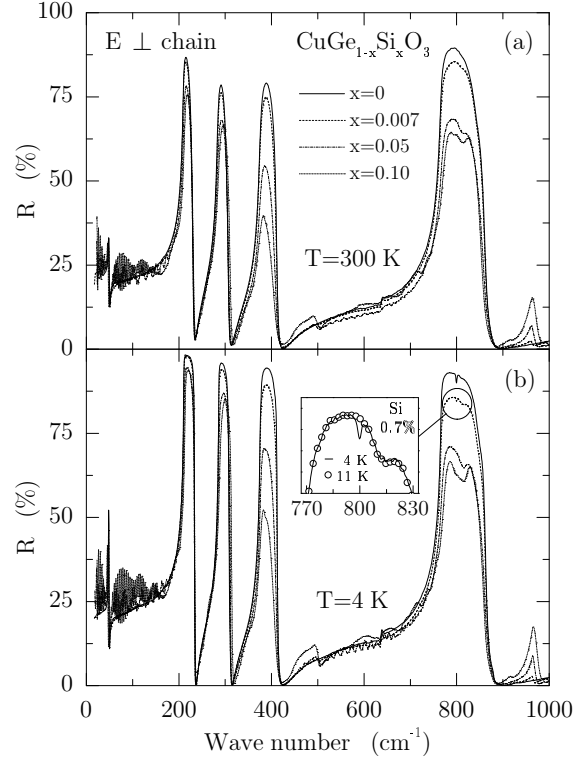,width=7.5cm,clip=}}
\vspace{.3cm}
\caption{$b$-axis reflectivity spectra of Si-doped single crystals of CuGeO$_3$,  
for different silicon concentrations, at two different temperatures: $T\!=\!300$ K (a), and 
$T\!=\!4$ K (b). 
The 800 cm$^{-1}$ folded mode, activated by the SP transition, is still observable for 0.7\% 
Si-doping by comparing (see inset) the 4 K (solid line) and the 11 K (circles) data.}
\label{cgdref2}
\end{figure}
\begin{figure}[t]
\centerline{\epsfig{figure=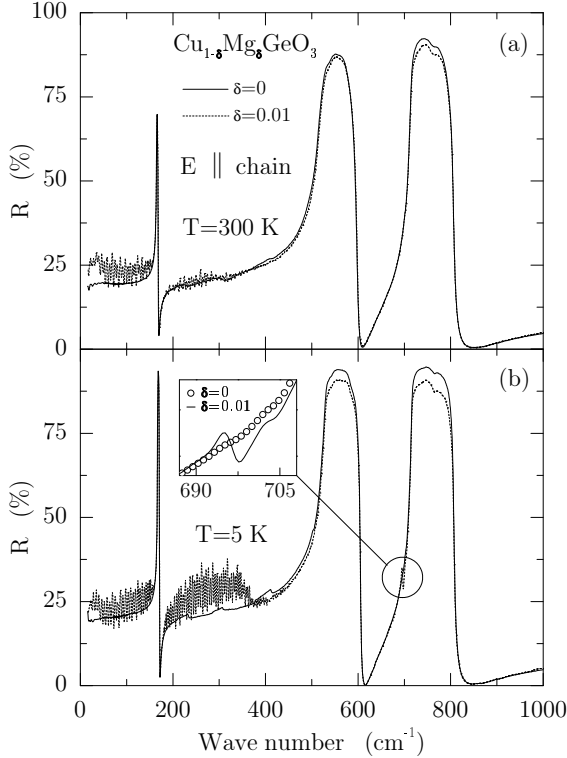,width=7.5cm,clip=}}
\vspace{.3cm}
\caption{$c$-axis reflectivity spectra of pure and 1\% Mg-doped single crystals of CuGeO$_3$,  
at two different temperatures: $T\!=\!300$ K (a) and $T\!=\!5$ K (b). 
The inset shows an enlarged view of the frequency region around 700 cm$^{-1}$ for  the 
data obtained on the pure (circles) and Mg-doped (solid line) samples, at $T\!=\!5$ K. 
The additional peak observed for Mg doping is due to the mass difference between Cu and Mg and 
it is not related to the SP transition.}
\label{cgdref3}
\end{figure}
\begin{figure}[t]
\centerline{\epsfig{figure=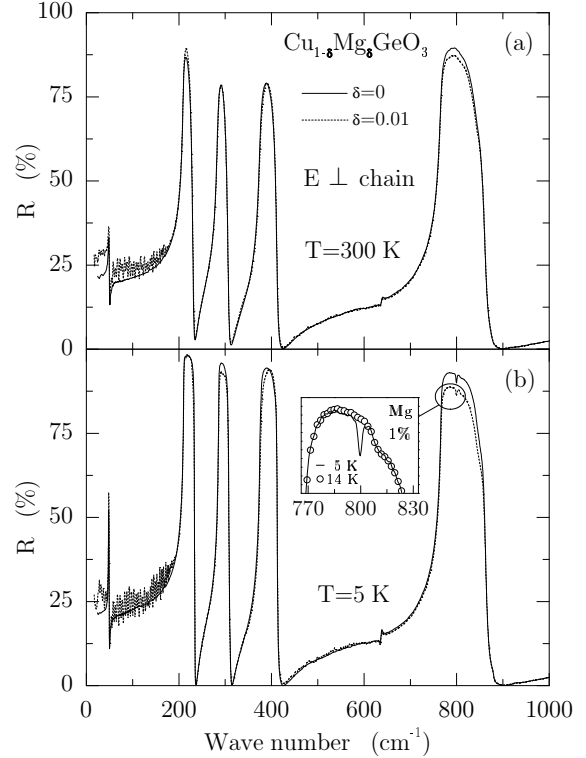,width=7.5cm,clip=}}
\vspace{.3cm}
\caption{$b$-axis reflectivity spectra of pure and 1\% Mg-doped single crystals of CuGeO$_3$, 
at two different temperatures: $T\!=\!300$ K (a) and $T\!=\!5$ K (b). For the Mg-doped sample the 
800 cm$^{-1}$ folded mode, activated by the SP transition,
is clearly observable  in the inset where the 5 K (solid line) and 
the 14 K (circles) data are presented.}
\label{cgdref4}
\end{figure}
\begin{figure}[t]
\centerline{\epsfig{figure=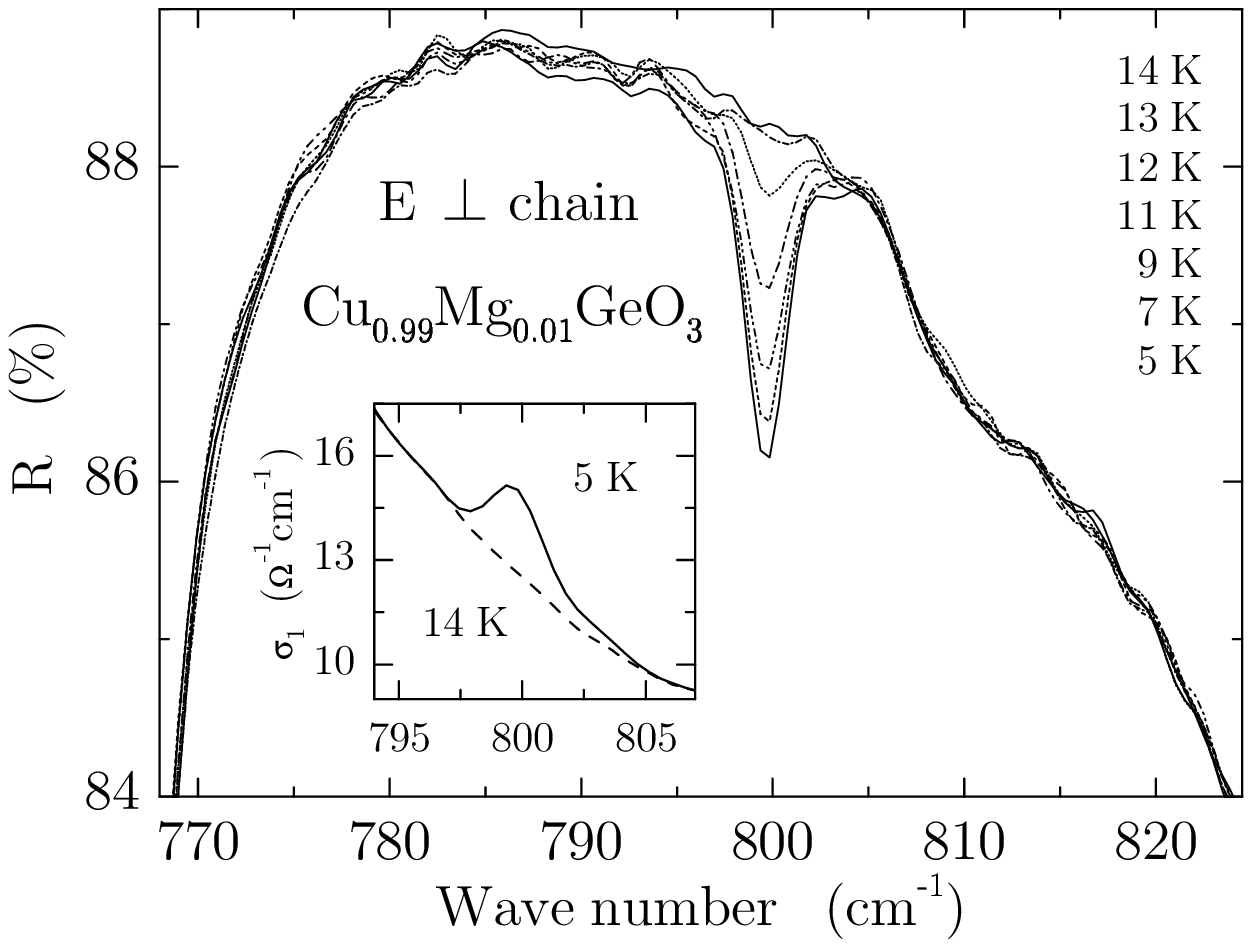,width=7.5cm,clip=}}
\vspace{.3cm}
\caption{Detailed temperature dependence of the 800 cm$^{-1}$ activated mode observed with 
$E\!\perp$\,chain in the reflectivity  spectra of Cu$_{\,0.99}$Mg$_{\,0.01}$GeO$_3$, for 
$T\!<\!T_{\rm{SP}}$. In the inset, where the dynamical conductivity calculated via Kramers-Kronig 
analysis is plotted, a new peak is clearly visible for $T\!=\!5$ K.}
\label{cgdsp1}
\end{figure}
\begin{figure}[t]
\centerline{\epsfig{figure=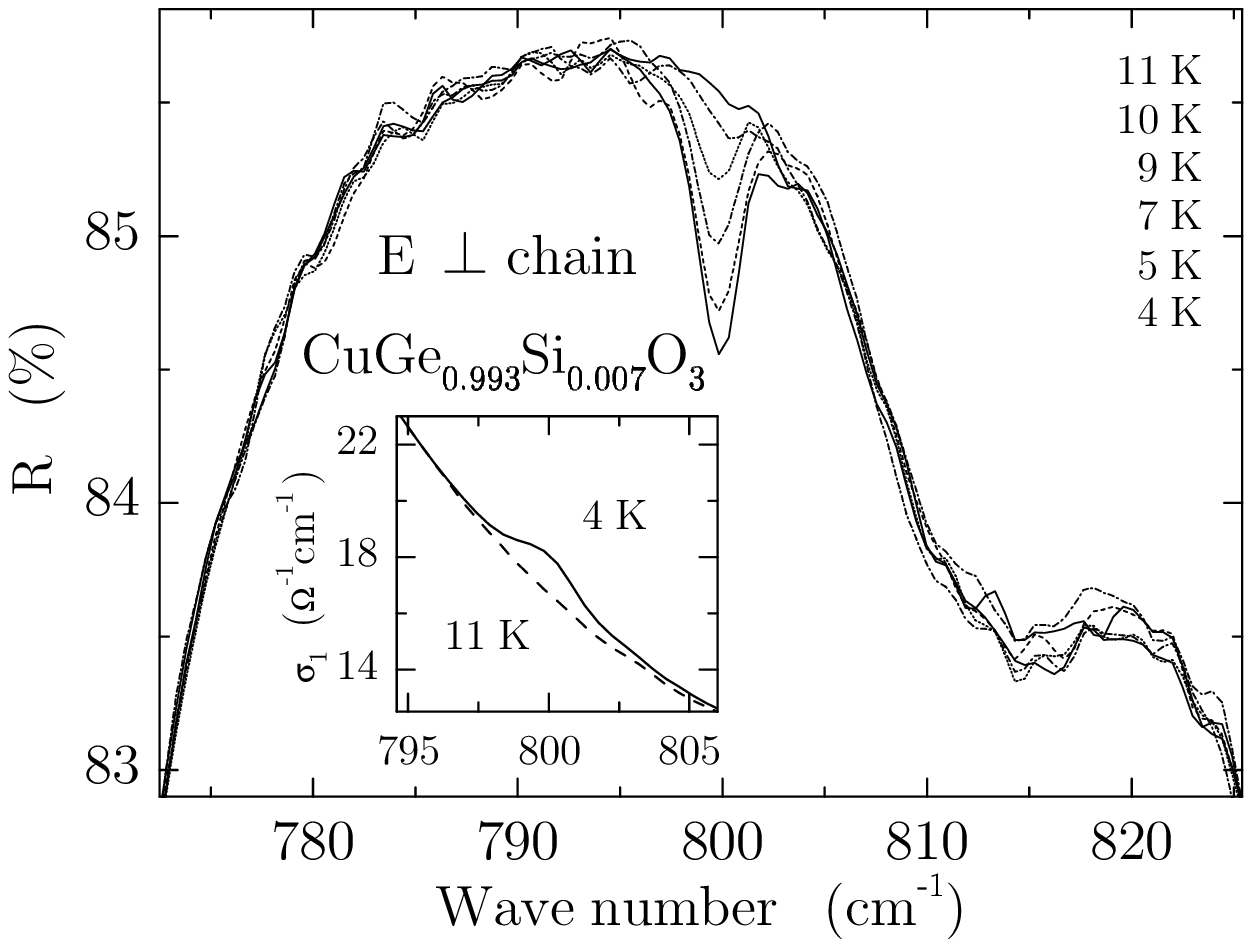,width=7.5cm,clip=}}
\vspace{.3cm}
\caption{Detailed temperature dependence of the 800 cm$^{-1}$ activated mode observed with 
$E\!\perp$\,chain in the reflectivity  spectra of CuGe$_{\,0.993}$Si$_{\,0.007}$O$_3$, for 
$T\!<\!T_{\rm{SP}}$. In the inset, where the dynamical conductivity calculated via Kramers-Kronig 
analysis is plotted, a new peak is clearly visible for $T\!=\!4$ K.}
\label{cgdsp2}
\end{figure}
\begin{figure}[t]
\centerline{\epsfig{figure=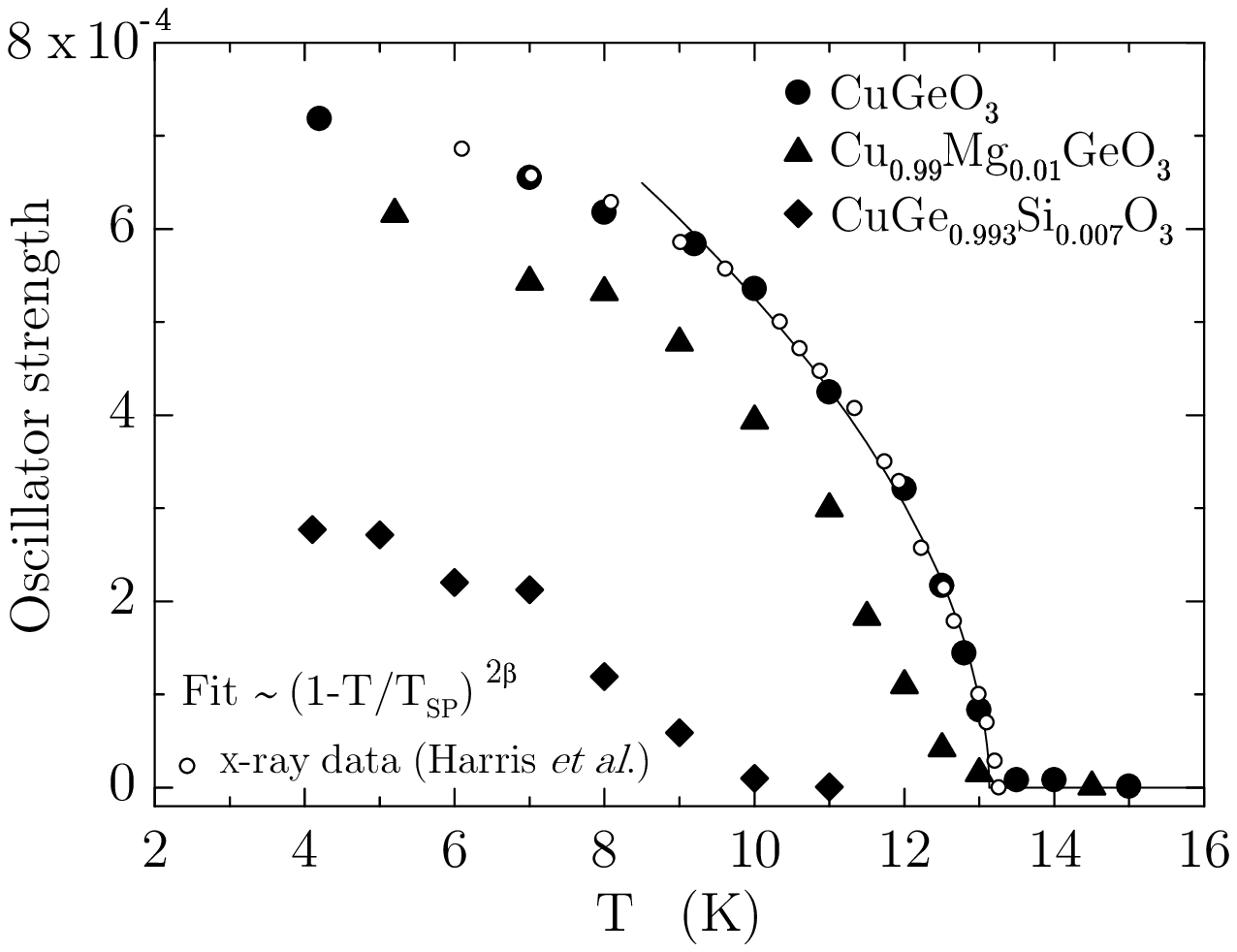,width=7.5cm,clip=}}
\vspace{.3cm}
\caption{Temperature dependence of the oscillator strength of  the zone boundary folded mode 
observed, for $E\!\perp$\,chain, at 800 cm$^{-1}$ on pure, 1\% Mg-doped, and 0.7\% Si-doped  
CuGeO$_3$ single crystals. The x-ray scattering data of Harris et al. \protect{\cite{harris}} 
and the fit to the power law  described in section\ \protect{\ref{cgphdist}} are also plotted, 
for pure CuGeO$_3$.}
\label{cgdsp3}
\end{figure}

\section{Doped CuGeO$_3$}
\label{cgd}

In this section we will present the optical spectroscopy data of several 
Cu$_{1-\delta}$Mg$_{\delta}$GeO$_3$  (with $\delta$\,=\,0,\,0.01), and  
CuGe$_{1-x}${\em B}$_{x}$O$_3$ [with {\em B}=Si ({\em x}\,=\,0,\,0.007,\,0.05,\,0.1), and Al 
({\em x}\,=\,0,\,0.01)] single crystals. In particular, we investigated, as we did for the 
pure material, the phonon spectrum as a function of temperature of Mg and Si substituted CuGeO$_3$, by 
means of reflectivity measurements in the far-infrared region (see section\ \ref{cgfirref}). 
The aim of this investigation was, first of all, to study the effect of doping on the SP 
phase transition; second, to verify the recent claims of Yamada and co-workers\cite{yamada,hidaka} 
who suggested that the structure  originally proposed for CuGeO$_3$, in the high temperature undistorted 
phase,\cite{vollenkle} could be wrong (see section\ \ref{cgfirref}). Moreover, transmission 
measurements in the mid-infrared region were performed on all the samples, in order to detect 
possible electronic and/or magnetic excitation which could provide us with additional 
information about the interplay of spin and charge in this quasi-1D systems 
(see section\ \ref{cgmirtra}).

\subsection{Far-Infrared Reflection}
\label{cgfirref}

The reflectivity data acquired on the Si doped samples for $E\!\parallel$\,chain ({\em c} axis), 
and $E\!\perp$\,chain ({\em b} axis), are shown in Fig.\ \ref{cgdref1} and\ \ref{cgdref2}, 
respectively. The spectra are similar to those we already discussed for pure CuGeO$_3$. However, 
some new features, stronger in intensity the higher the Si concentration, are observable already 
at room temperature. Therefore, they are due to the substitution of Ge with the lighter Si and 
not directly related to the SP transition: New phonon peaks at 900 cm$^{-1}$, along the {\em c}  
axis (Fig.\ \ref{cgdref1}), and at 500 and 960 cm$^{-1}$, along the {\em b} axis 
(Fig.\ \ref{cgdref2}). Moreover, the much more complicated line shape and the considerable 
reduction of the oscillator strength of the high frequency phonons indicate: First, a strong 
Ge (Si) contribution to these modes, mainly due to O vibrations.\cite{popovic} Second, that  
substituting Si for Ge has apparently a strong symmetry-lowering effect, a point which we will 
discuss again later in relation to the claim of Yamada {\em et al.}\cite{yamada,hidaka} about a 
possible lower symmetry, with respect to the one originally proposed by V\"ollenkle 
{\em et al.},\cite{vollenkle} for the uniform structure of CuGeO$_3$. At temperatures 
lower than $T_{\rm{SP}}$ (Fig.\ \ref{cgdref2}b), we can observe the folded mode at 
800 cm$^{-1}$ along the {\em b} axis, previously detected on pure 
 CuGeO$_3$ (see section\ \ref{cgfirref}). However, it is found only for the lowest Si concentration, 
as shown in the inset of 
Fig.\ \ref{cgdref2}b, where the 4 K and 11 K data are compared. We can conclude that up to 0.7\% 
Si doping the SP transition is still present with $T_{\rm{SP}}\!<$11 K, whereas for 5\% and 10\% 
Si concentrations no signature of the transition could be found in our spectra. 

The results obtained on the 1\% Mg-doped sample are plotted, together with the results obtained 
on pure CuGeO$_3$, in Fig.\ \ref{cgdref3} and Fig.\ \ref{cgdref4} for $E\!\parallel$\,chain 
({\em c} axis), and $E\!\perp$\,chain ({\em b} axis), respectively. Clearly, Mg doping is 
affecting the optical response of  CuGeO$_3$  less than Si doping does. A new phonon, due 
to the mass difference between Cu and Mg, is present in the {\em c}-axis spectra at 695 
cm$^{-1}$, as clearly shown in the inset of  Fig.\ \ref{cgdref3}b, for $T\!=\!5$ K. Moreover, we 
clearly observe for $E\!\perp$\,chain (see inset of  Fig.\ \ref{cgdref4}b),  the 800 cm$^{-1}$  
zone boundary mode activated by the SP transition. On the one hand, for the 1\% Mg-doped sample, 
$T_{\rm{SP}}$ seems to be lower than in pure CuGeO$_3$; on the other hand, the structural 
deformation is not as strongly reduced as in the 0.7\% Si-doped sample, as can be deduced 
from the direct comparison between  the insets of  Fig.\ \ref{cgdref2}b and Fig.\ \ref{cgdref4}b. 

In order to estimate more precisely the reduction the oscillator strength for the 800 
cm$^{-1}$ phonon, and eventually the decrease of $T_{\rm{SP}}$ with respect to what was 
observed on pure CuGeO$_3$, 
we made careful measurements for $T\!<\!T_{\rm{SP}}$ on the 1\% Mg and the 0.7\% 
Si-doped samples (see Fig.\ \ref{cgdsp1} and\ \ref{cgdsp2}, respectively). In both cases we can 
observe, in the reflectivity spectra, the gradual disappearance of the activated phonon for 
$T\!\rightarrow\!T_{\rm{SP}}$ from below and, in the optical conductivity (see insets of 
Fig.\ \ref{cgdsp1} and\ \ref{cgdsp2}), the additional peak present in the dimerized 
phase data, superimposed 
to the background of the $B_{2u}$ phonon of the high symmetry phase. As for the pure material 
(section\ \ref{cgphdist}), by fitting the reflectivity spectra with Lorentz oscillators for 
the optical phonons, we could obtain the temperature dependence of the oscillator strength for 
the 800 cm$^{-1}$ line. The results are plotted for all the samples in Fig.\ \ref{cgdsp3}. 
For pure  CuGeO$_3$ we saw that our results for the oscillator strength of the activated optical phonon 
are in very good agreement with the peak intensity of a superlattice reflection measured by 
Harris {\em et al.}\cite{harris} (see Fig.\ \ref{cgdsp3}). Similarly, the data obtained on doped 
samples,  compare nicely to those presented, for 0.7\% Si 
doping, in Ref.\ \onlinecite{regnault} and, for 1\% Mg doping, to those reported in 
Ref.\ \onlinecite{sasago} for an 0.9\% Zn-doped sample (superlattice reflection data for Mg-doped 
CuGeO$_3$, obtained by neutron-scattering,  became available in the literature only very 
recently,\cite{nakao} however for a too high Mg concentration to be directly comparable to our 
results).  

It is not possible to perform the same fit for the data acquired on doped samples, as 
we did on the pure crystal, 
because they are characterized by an upturned curvature near $T_{\rm{SP}}$, which can be explained 
in terms of a distribution of transition temperatures due to the disorder introduced upon 
doping the system.\cite{regnault} However, for the 1\% Mg and the 0.7\% Si doped samples 
the estimates for $T_{\rm{SP}}$ of approximately 12.4 K and 9.3 K, respectively, can be obtained  
(Fig.\ \ref{cgdsp3}). Considering that usually, for pure CuGeO$_3$, $T_{\rm{SP}}\!\approx\!14.2$ 
(on our pure sample a slightly reduced value has been observed), we can conclude 
that Si is three times more effective in reducing $T_{\rm{SP}}$ than Mg. This result has recently 
been confirmed, on a more strong basis,  by a very systematic and detailed investigation of the 
decrease of $T_{\rm{SP}}$ and of the occurrence of 3D AF order at lower temperature on several 
doped single crystals of CuGeO$_3$, by magnetic susceptibility measurements.\cite{grenier} 

The difference between Mg and Si in influencing the phase transition can be understood 
considering what is the effect of these two different ways of doping on the magnetism of the system. 
The substitution of a Cu$^{2+}$ ion by a non magnetic impurity, like Mg, effectively cuts the 
CuO$_2$ chain in two segments and, at the same time,  breaks a singlet dimer. Therefore, a free 
S=1/2 Cu$^{2+}$ spin is created for each Mg impurity. More subtle is the effect of the 
substitution of Ge with Si: As shown by Khomskii and co-workers,\cite{geertsma,khomskii} it has 
to do with the question why the Cu-Cu nn superexchange interaction in  CuGeO$_3$ is 
AF in the first place. In fact, because the Cu-O(2)-Cu bond angle is 
$\gamma\!\approx\!98^{\circ}$, and because the Goodenough-Kanamory-Anderson 
rule\cite{goodenough} states that the 90$^{\circ}$ superexchange of two magnetic ions 
(via a ligand), with partially filled {\em d} shells, is (weakly) ferromagnetic, it is not that 
obvious why the nn superexchange is AF in CuGeO$_3$. Khomskii and 
co-workers\cite{geertsma,khomskii} calculated that the 8$^{\circ}$ deviation from the critical 
value of 90$^{\circ}$ is not sufficient for the superexchange to change sign. The AF 
interaction was shown to be a 
consequence of the side-group effect. In fact, because of the presence of Ge and, in particular, 
because of the Ge-O hybridization, the oxygen $p_x$ and $p_y$ orbitals are not equivalent any more. 
Therefore, the cancellation of the AF contributions to the Cu-Cu superexchange, 
via the ligand O, is no longer complete, resulting in a (weak) AF superexchange. 
When Ge is substituted with Si, one could at first glance expect a relatively weak effect because, 
contrary to the substitution of Cu with Mg, Si is not directly breaking a dimer.
However, because Si is smaller than Ge, both the Si-O bond length [which is proportional  
the Cu-O(2)-Cu bond angle $\gamma$] and the Si-O hybridization are reduced.\cite{geertsma,khomskii} 
As a result, Si breaks the Cu-Cu superexchange interaction on the two neighboring CuO$_2$ chains 
adjacent to the Si$^{4+}$ ion. This explains, at least qualitatively, why Si has to be much more 
efficient than Mg in reducing $T_{\rm{SP}}$. 
\begin{figure}[t]
\centerline{\epsfig{figure=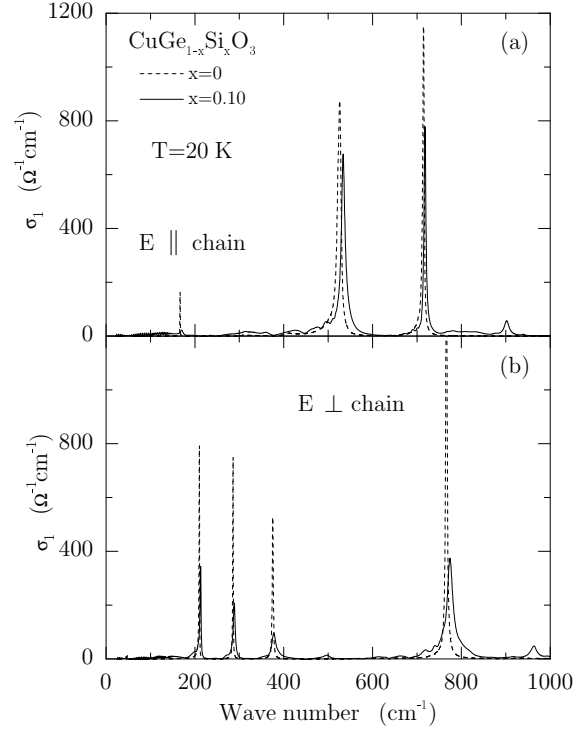,width=7.5cm,clip=}}
\vspace{.3cm}
\caption{Low temperature optical conductivity spectra, obtained by Kramers-Kronig analysis, 
of pure and 10\% Si-doped CuGeO$_3$, for 
$E\!\parallel$\,chain (a) and $E\!\perp$\,chain (b). Note that in panel (b) the data on 
pure  CuGeO$_3$ have ben clipped in order to use the same scale as in panel (a): in fact, the peak 
value of the phonon at 766 cm$^{-1}$ is 2010 $\Omega^{-1}$cm$^{-1}$.}
\label{cgdcon1}
\end{figure}
\begin{figure}[t]
\centerline{\epsfig{figure=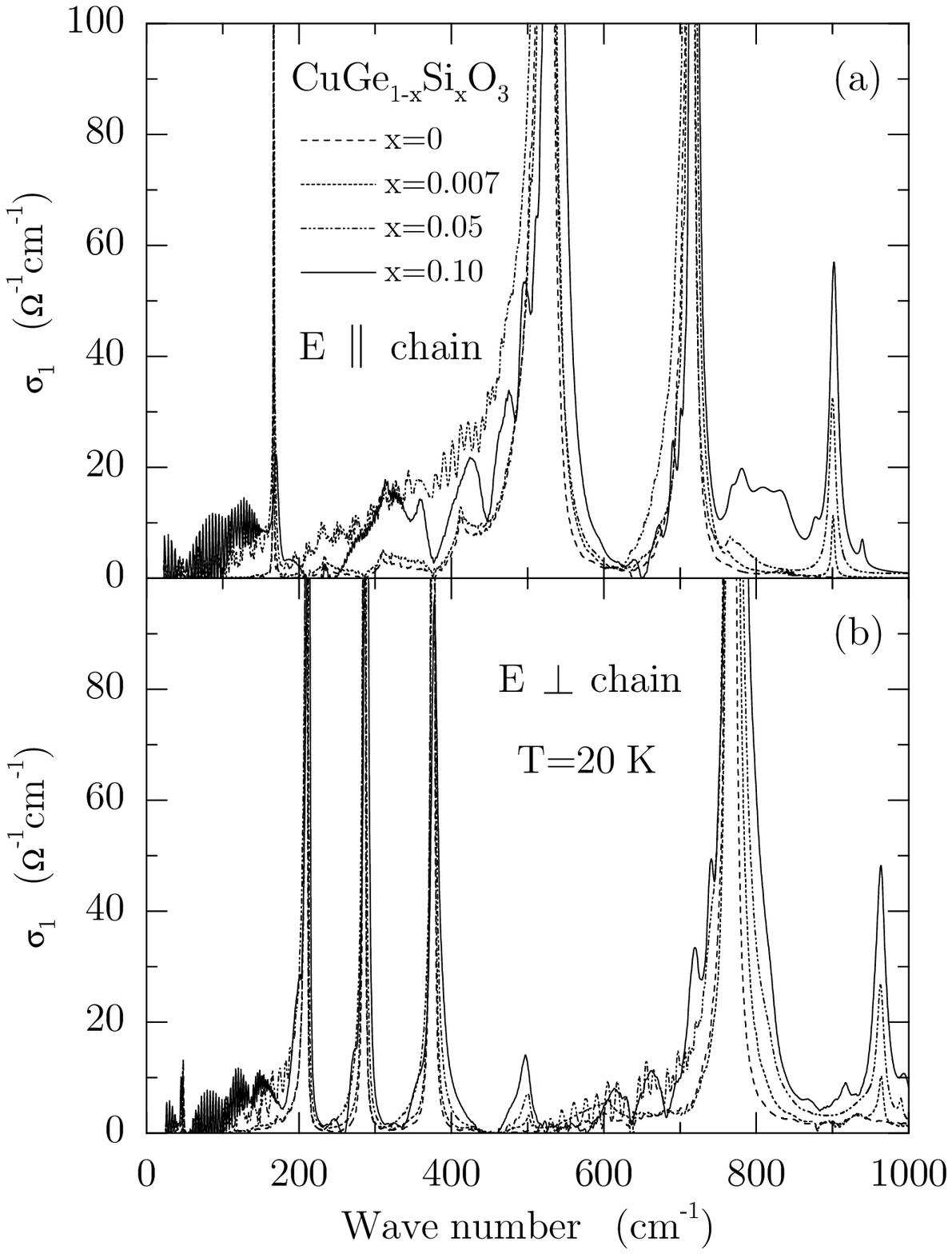,width=7.5cm,clip=}}
\vspace{.3cm}
\caption{Low temperature optical conductivity, obtained by Kramers-Kronig analysis, 
of pure and several Si-doped  CuGeO$_3$ single crystals, for 
$E\!\parallel$\,chain (a) and $E\!\perp$\,chain (b).}
\label{cgdcon2}
\end{figure}

Recently Yamada and co-workers\cite{yamada,hidaka} raised doubts about  the structure of  
CuGeO$_3$ in the high-temperature 
undistorted phase. They showed that the results of electron paramagnetic resonance 
experiments\cite{yamada} can be fully understood only taking into account spin-antisymmetric 
interaction, such as the Dzyaloshinsky-Moriya exchange interaction.\cite{moriya} However, 
the antisymmetric exchange interaction is  strictly forbidden in the space group {\em Pbmm} because 
the midpoint in between two adjacent Cu$^{2+}$ ions, along the CuO$_2$ chains, has inversion 
symmetry.  To verify these findings, x-ray diffraction experiments were performed by Hidaka 
{\em et al.},\cite{hidaka} at room temperature, on pure single crystals grown by the floating-zone 
method, subsequently improved in their quality by a combined annealing and slow cooling process. 
As a result, new superlattice reflections were found, related to tilting and rotation of the 
GeO$_4$ tetrahedra in four different manners along the {\em c} axis, and in antiphase along the 
{\em a} axis. As a consequence, local distortions of the CuO$_6$ octahedra are induced, with a 
four times periodicity along the {\em c} axis. The space group was determined to be $P2_12_12_1$, 
with an eight times larger unit cell, i.e., $2a\!\times\!b\!\times\!4c$, with respect to the one 
originally proposed by V\"ollenkle {\em et al.}.\cite{vollenkle} According to Hidaka 
{\em et al.},\cite{hidaka} this additional deformation of the structure would be there also on non 
carefully annealed single crystals, but only on a short-range scale. Moreover, the newly proposed 
structure allows for the antisymmetric exchange interaction which was required to interpret the 
electron paramagnetic resonance results\cite{yamada} on both annealed and as-grown samples, 
because the Dzyaloshinsky-Moriya exchange interaction depends only on the local symmetry around the 
nn spins.\cite{moriya} 

This could have strong consequences for the  
picture of the SP transition in CuGeO$_3$, because it implies the presence of four different 
values of the superexchange interaction at room temperature and, upon decreasing 
the temperature toward $T_{\rm{SP}}$, strong fluctuations of the exchange interactions, induced by 
the increasing lattice instability. As the $P2_12_12_1$ group has fewer symmetry elements than 
the conventional {\em Pbmm}, a larger number of phonon modes should be observed in the optical 
spectra. Skipping the details of the calculation, the irreducible representations of the optical 
vibrations of  CuGeO$_3$ in the high temperature phase, for the space group $P2_12_12_1$, is:
\begin{eqnarray}
  \Gamma^{\prime\prime}\!=&&\!56A(aa,bb,cc)\!+\!55B_{1}(ab;E\|c)\nonumber\\
                           &&+63B_{2}(ac;E\|b)\!+\!63B_{3}(bc;E\|a) \, ,            
\label{cgrepY}
\end{eqnarray} 
corresponding to 63, 63, and 55 optical active modes along the {\em a}, {\em b}, and {\em c} axes, 
respectively. 
As we already discuss in section\ \ref{cgphdist}, the number of phonons detected in the 
reflectivity spectrum of pure  CuGeO$_3$ was in agreement with the results of the group theoretical 
analysis for the space group {\em Pbmm} proposed by V\"ollenkle {\em et al.}\cite{vollenkle} 
At the same time, as observed by Yamada and co-workers,\cite{yamada,hidaka} the additional 
distortion, characterizing the $P2_12_12_1$ space group, is present only on a short-range scale in 
as-grown samples, like the ones used in this experimental work. However, even on pure CuGeO$_3$, in 
fitting the reflectivity spectra with Lorentz oscillators for the optical phonons, we had to add, 
in order to be able to exactly reproduce the experimental data, a number of very weak and broad  
additional peaks, of unclear origin. In this context, the results obtained on Si substituted  
CuGeO$_3$ single crystals seem to be more significant. We already mentioned how strongly the 
phonon spectra were affected by the presence of Si in the samples. 
This effect is even more evident when the optical conductivity spectra are considered. In 
Fig.\ \ref{cgdcon1}, the low temperature optical conductivity of pure and 10\% Si-doped 
CuGeO$_3$, for $E\!\parallel$\,chain (a) and $E\!\perp$\,chain (b), is plotted. We can observe a 
reduction of oscillator strength and a broadening for all the lattice vibrations. Moreover, 
additional bands appear over the all range, as clearly shown in 
Fig.\ \ref{cgdcon2}, where the low temperature optical conductivity, for all the measured Si-doped 
samples, is presented. We can see that not only sharp peaks appear at  900 and 960 
cm$^{-1}$ along the {\em c} and {\em b} axes, respectively, as discussed at the beginning of 
this section.  There is also a large number of phonon bands, growing in intensity upon 
increasing the Si concentration, at both low and high frequencies. It is obviously not possible  
to check in detail whether we detected all the modes expected, on the basis of group 
theory, for the newly proposed space group $P2_12_12_1$:\cite{yamada,hidaka} There are simply 
too many of them, in both theory (see Eq.\ \ref{cgrepY}) and experiment (see Fig.\ \ref{cgdcon2}). 
Moreover, we have to take into account that, in general, the introduction of impurities in a 
crystal results in a relaxation of the {\em k} conservation rule. Therefore, states with 
$k\!\neq\!0$ are projected back to the $\Gamma$ point, and phonon bands averaged over the entire 
Brillouin zone can be measured in an optical experiment. However, we may 
speculate that Si, that has such a strong influence on $T_{\rm{SP}}$ via the side group 
effect,\cite{geertsma,khomskii} could also be responsible for enhancing the underlying lattice 
distortion which was suggested to be already present in pure CuGeO$_3$, 
by Yamada and co-workers.\cite{yamada,hidaka} 
In our opinion these results, although not conclusive, indicate that a symmetry lower than 
the one assumed until now could be possible. In particular, it should be taken into 
account as one of the possible explanations for the aspects still unresolved of the physics of 
 CuGeO$_3$ (see section\ \ref{cgdisc}).

\subsection{Mid-Infrared Transmission}
\label{cgmirtra}
\begin{figure}[t]
\centerline{\epsfig{figure=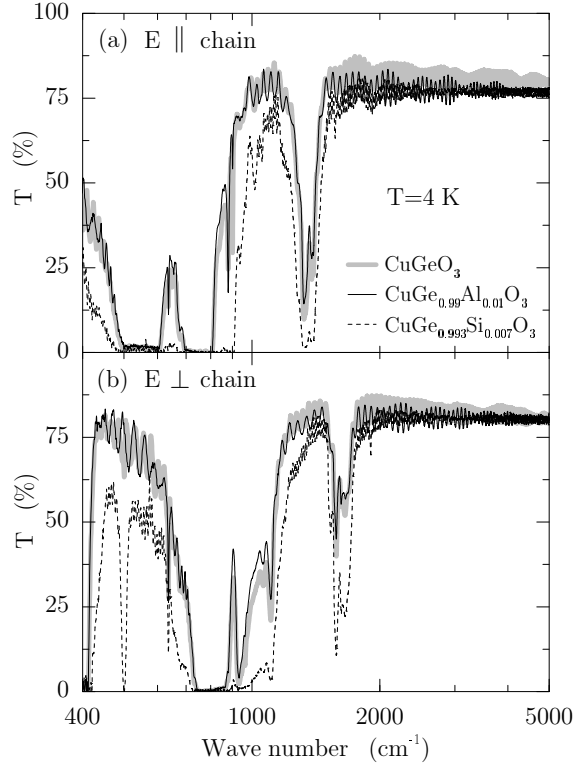,width=7.5cm,clip=}}
\vspace{.3cm}
\caption{Low temperature transmission spectra of pure, 1\% Al, and 0.7\% Si doped CuGeO$_3$, for 
$E\!\parallel$\,chain (a) and $E\!\perp$\,chain (b).}
\label{cgdts1}
\end{figure}
\begin{figure}[t]
\centerline{\epsfig{figure=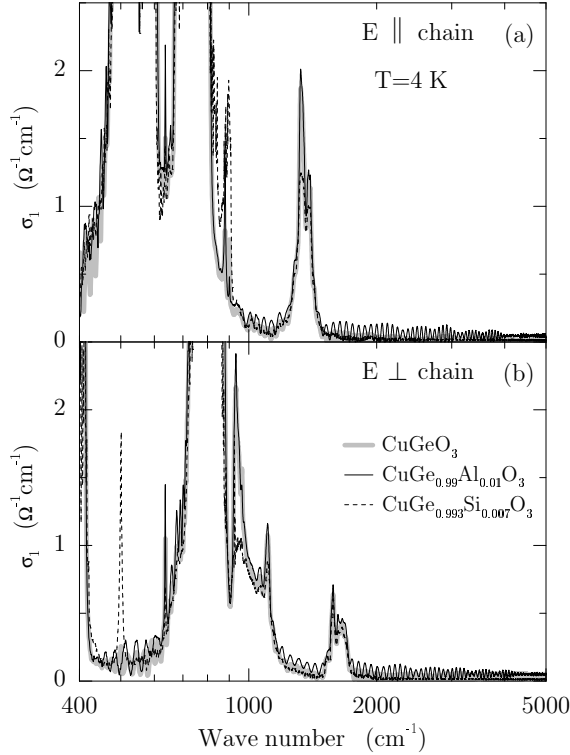,width=7.5cm,clip=}}
\vspace{.3cm}
\caption{Low temperature conductivity spectra, obtained by direct inversion of the Fresnel 
formula, of pure, 1\% Al, and 0.7\% Si doped CuGeO$_3$, for $E\!\parallel$\,chain (a) and 
$E\!\perp$\,chain (b). Note the very low value of $\sigma_1(\omega)$.}
\label{cgdts2}
\end{figure}
In this section we will discuss the optical data obtained by performing transmission experiments, 
in the range going from 400 to 8000 cm$^{-1}$, on several 
Cu$_{1-\delta}$Mg$_{\delta}$GeO$_3$  (with $\delta$\,=\,0,\,0.01), and  
CuGe$_{1-x}${\em B}$_{x}$O$_3$ [with {\em B}=Si ({\em x}\,=\,0,\,0.007,\,0.05,\,0.1), and Al 
({\em x}\,=\,0,\,0.01)] single crystals. In particular, by combining reflectivity and transmission 
results, we could calculate the optical conductivity by direct inversion of the Fresnel 
formula.\cite{klein} By this investigation, we aimed to study whether, in CuGeO$_3$, doping has an  
effect on the electronic and/or magnetic excitations similar to what has been observed on the 
underdoped 
parent compounds of high-$T_c$ superconductors.\cite{kastner,markus,markus1} In fact, on the latter 
materials, chemical substitution introduces charge carriers in the 2D Cu-O planes, resulting in a 
disappearance of the gap and in the transfer of spectral weight into a Drude peak, for the in-plane 
optical response. Moreover, the so called `mid-infrared' band, whose interpretation is still 
controversial, is usually present in these systems.\cite{kastner,markus,markus1} 

The low temperature (4 K) transmission spectra of pure, 
1\% Al, and 0.7\% Si doped CuGeO$_3$, for $E\!\parallel$\,chain and $E\!\perp$\,chain, are 
plotted in Fig.\ \ref{cgdts1}a and \ref{cgdts1}b, respectively. Spectra at different temperatures 
above and below the phase transition were measured. However, the results are not shown because 
no particular temperature dependence was observed in cooling down the sample from room temperature 
to $T_{\rm{SP}}$, nor across the phase transition. Similarly, the spectra for 1\% Mg and 5 and 10\% Si 
doped crystals are not presented because no additional information can be obtained from these 
results. As shown by Fig.\ \ref{cgdts1}, the transmission through pure CuGeO$_3$, in this frequency 
region, is mainly  characterized by the strong absorptions of the phonons of the high symmetry 
phase at 528 and 715 cm$^{-1}$, along the {\em c} axis, and at 376 and 766 cm$^{-1}$, along the 
{\em b} axis. In addition, many multiphonon bands are present, the most intense ones being at 
1330 and 1580 cm$^{-1}$ along the {\em c} and {\em b} axes, respectively. The same features are 
observable on the doped samples where, however, more multiphonon bands are detected reflecting the 
presence of additional peaks in the single-phonon spectrum, already discuss in 
section\ \ref{cgfirref} on the basis of reflectivity measurements. A last remark has to be made 
about the oscillations present in all the spectra of Fig.\ \ref{cgdts1}. These are, as 
mentioned before, interference fringes due to Fabry-Perot resonances;\cite{klein} the 
different period of the interference pattern for the three crystals is simply due to the different 
thickness of the samples used in the experiments. 

If we now consider the optical conductivity 
spectra plotted in Fig.\ \ref{cgdts2}, it is clear that nothing else has been detected, in this 
frequency range, besides absorption processes purely related to lattice degrees of freedom. In 
particular, for all the sample, the conductivity is  just zero from 2000 to 8000 cm$^{-1}$. From 
these results and from the fact that no signature of a Drude peak was observed in any of the doped 
samples in reflectivity experiments (see section\ \ref{cgfirref}), we conclude that no 
charge carriers are introduced in the  CuGeO$_3$ by the different chemical substitutions we tried, nor 
magnetic/dielectric polarons (observed, e.g., in ultra low doped  
YBa$_2$Cu$_3$O$_6$\,\,\cite{markus,markus1}). One has to note that, in this sense, the most 
significant of the results we showed is the one obtained on Al substituted   CuGeO$_3$ because in this 
case, contrary to Si substitution, Ge is replaced by an element with  different valence.

\section{Discussion}
\label{cgdisc}

As shown in the course of this paper, in our investigation of pure and doped  CuGeO$_3$ with 
optical spectroscopy, we observed features which are not completely understandable on the basis of 
Cross and Fisher theory for the SP phase transition,\cite{crossfi} and of the achieved picture 
of the SP transition in CuGeO$_3$. No pre-existing soft mode was detected, not only in our 
optical measurements, but especially in neutron scattering experiments.\cite{braden1} This result 
could explain why the value of $\beta$ in the expression describing the temperature dependence of 
the order parameter close to $T_{\rm{SP}}$ [$\delta\!\sim\!(1-T/T_{\rm{SP}})^{\beta}$] deviates, 
in CuGeO$_3$, from the mean-field behavior $\beta\!=\!1/4$ observed in the TTF salt (see 
section\ \ref{cgphdist}).
Moreover, a direct singlet-triplet 
excitation across the magnetic gap, possibly at the wave vector (0,2$\pi\!/b$,0) in the Brillouin 
zone, was detected in our transmission spectra. This excitation is, in principle, `doubly' 
forbidden: In fact, because of symmetry considerations and of spin conservation we are restricted, 
in an optical experiment, to excitations characterized by $k\!=\!0$ and $\Delta S\!=\!0$. Finally, 
strong changes in the phonon spectrum were observed on Si-substituted samples, which might be 
explained in terms of the alternative space group $P2_12_12_1$, recently proposed for  CuGeO$_3$ in the 
high temperature uniform phase.\cite{yamada,hidaka} 

Many attempts have recently been made to improve the understanding of the SP phase transition 
in CuGeO$_3$. In particular, it has been shown that the adiabatic treatment of the 3D phonon system, 
characteristic for the Cross and Fisher theory,\cite{crossfi} is not appropriate for the case of 
CuGeO$_3$, where the phonons contributing appreciably to the SP distortion  
have energies much higher that the magnetic gap.\cite{uhrig,gros} An alternative description 
of the SP transition, not based on the assumption of phononic adiabaticity, was developed by 
Uhrig:\cite{uhrig} Phonons are considered as the fast subsystem, responsible for the interchain 
coupling, and an effective dressed spin model is derived. As a result, the soft phonon is absent 
and the SP transition is characterized by growing domains of coherent dimerization, whose size 
diverges at $T_{\rm{SP}}$. Alternatively, Gros {\em et al.}\cite{gros} showed that no inconsistency 
is present in Cross and Fisher theory:\cite{crossfi} They concluded that, in Cross and Fisher 
framework, a soft phonon has to be present only if the bare phonon frequency satisfies the 
relation $\Omega_0\!<\!2.2\,T_{\rm{SP}}$. For larger phonon frequencies only a central peak is 
expected at $T_{\rm{SP}}$. However, this new collective excitation, which would consist of the 
linear superposition of a phonon with two magnons in a singlet state,\cite{gros} has not been 
observed up to now. Moreover, it has been shown that for a detailed understanding of magnetic 
susceptibility\cite{hase,pouget} and magnetostriction\cite{buchner,lorenz1} data and, at the 
same time, of the singlet and triplet excitation branches below the continuum,\cite{els,ain}  
not only the 2D character of the system cannot be neglected,\cite{nishi,uhrig} but also both 
the nn ($J_1$) and the nnn ($J_2$) magnetic exchange interactions have to be taken into 
account in  CuGeO$_3$.\cite{lorenz1,fabricius,bouzerar} The system would then be described by an 
alternating and frustrated AF 1D Heisenberg spin-chain model:
\begin{equation}
  H=J \sum_j \left\{ \left[1+\delta(-1)^i \right] \vec{S}_j\cdot\vec{S}_{j + 1} 
  +\alpha \vec{S}_j\cdot\vec{S}_{j + 2} \right\} \, , 
\label{cghsp}
\end{equation}
where $\delta$ is the static dimerization parameter and $\alpha\!=\!J_1/J_2$ is the frustration 
parameter. In particular, the value $\alpha\!=\!0.354$ was 
obtained,\cite{lorenz1,fabricius,bouzerar} 
which is significantly larger than the critical value sufficient for the formation of a 
spontaneous gap in the magnetic excitation spectrum, in absence of lattice dimerization, i.e., 
$\alpha_c\!=\!0.241$.\cite{castilla,okamoto} However, on the basis of this approach and, in 
particular, with the large value obtained for $\alpha$, the amplitude of the dimerization $\delta$, 
estimated by reproducing the singlet-triplet excitation gap,\cite{riera,bouzerar1} is substantially 
underestimated. In fact in this way, the very small value $\delta\!\simeq\!0.012$ is obtained, 
whereas, from the combined analysis of the structural lattice distortion in the dimerized phase 
and of the magneto-elastic coupling in the uniform phase (from magnetostriction measurements), 
the value $\delta\!\simeq\!0.04\!-\!0.05$ results.\cite{buco} Still retaining the value 
$\alpha\!=\!0.354$, a consistent value for the lattice dimerization (i.e., of the order of 5\%) 
is obtained from the analysis of the inelastic neutron scattering data if, instead of a static 
dimerization parameter $\delta$, an explicit coupling between the spins and the three dimensional 
phonon system is introduced in the Hamiltonian given in Eq.\ \ref{cghsp}.\cite{wellein} 

In relation to the space group $P2_12_12_1$, recently proposed for  CuGeO$_3$ in the 
high temperature phase,\cite{yamada,hidaka} besides the optical data on Si-substituted 
samples we presented, very interesting are the inelastic neutron scattering results reported by 
Lorenzo {\em et al.}\cite{lorenzo} A second singlet-triplet (optical) branch was observed in pure 
CuGeO$_3$, with a gap value at the $\Gamma$ point of $\sim\!5.8$ meV.
The dispersion of this newly found optical mode is identical to the one of the acoustic mode, 
but shifted by ($2\pi/b+\pi/c$). Lorenzo {\em et al.}\cite{lorenzo} proposed that the 
origin of this second mode is the relative orientation of the Cu-O(2)-Cu units between next 
neighboring chains along the $b$ axis, along with a small spin-orbit coupling. This would give rise  
the slight distortion of the spin isotropy necessary to reproduce the finite intensity for the 
optical triplet mode and the difference in the scattering intensities between the optical and the 
acoustic branches.\cite{lorenzo}  On the other hand, anisotropy in the magnetic exchange 
constants along the $b$ axis, below $T_{\rm{SP}}$, ought to be present in the space group 
$P2_12_12_1$ proposed by Yamada {\em et al.},\cite{yamada,hidaka} because of the strong 
fluctuations of the exchange interactions expected upon decreasing the temperature toward 
$T_{\rm{SP}}$.

These results are not only showing that the structure generally assumed for  CuGeO$_3$  in the 
uniform phase may be incorrect, but they are also suggesting a possible explanation for the 
singlet-triplet excitation we observed in transmission at $\sim\!44$ cm$^{-1}$\,$\sim\!5.5$ meV. 
In fact, such an optical  transition, with approximately the right energy value, is now present 
at $k\!=\!0$ and not only 
at the wave vector (0,2$\pi\!/b$,0), which is not accessible in an optical experiment. The final 
constraint $\Delta S\!=\!0$  which is not satisfied for this excitation, could be possibly 
explained as a consequence of spin-orbit interaction, which  relaxes the spin conservation rule. 
In order to assess whether this absorption line is given to an electric dipole or to a magnetic 
dipole transition, in presence of spin-orbit interaction, a careful analysis of selection 
rules and absolute intensities is needed. 

\section{Conclusions}
\label{cgconc}

In this paper we have been discussing in detail the temperature dependent optical response of 
pure and doped CuGeO$_3$, in the frequency range going from 20 to 32\,000 cm$^{-1}$, with 
particular emphasis on the infrared phonon spectra. We could 
detect zone boundary folded modes activated by the SP phase transition. 
Following the temperature dependence of these modes we were able to determine the second order 
character of the phase transition and to study the effect of doping on $T_{\rm{SP}}$: 
In particular, we showed that the substitution of Ge with Si is three times more efficient, 
than the one of Cu with Mg, in reducing $T_{\rm{SP}}$. This was explained, following Khomskii and 
co-workers,\cite{geertsma,khomskii} as a consequence of the side-group effect. 

Moreover, in transmission experiments we detected a direct singlet-triplet excitation, across 
the magnetic gap, which is not understandable on the basis of the magnetic excitation spectrum  
generally assumed for  CuGeO$_3$.  The optical activity of this excitation has been discussed in 
relation to newly reported inelastic neutron scattering data which show the existence of a  
second (optical) magnetic branch.\cite{lorenzo} The anisotropy in the magnetic exchange 
constants along the $b$ axis, necessary for the optical triplet mode in order to gain a 
finite intensity, and the 
strong changes in the phonon spectra of Si-substituted samples might be explained in terms of 
the alternative space group $P2_12_12_1$, recently proposed for  CuGeO$_3$ in the high temperature 
uniform phase.\cite{yamada,hidaka}

\section{Acknowledgements}

We gratefully acknowledge M. Mostovoy and D.I. Khomskii for stimulating discussions, and P.H.M. 
van Loosdrecht, J.E. Lorenzo, and  M. Gr\"uninger  for many useful comments. One of us (A.D.) 
is pleased to thank M. Picchietto and B. Top$\acute{\rm{\i}}$ for assistance. This investigation was 
supported by the Netherlands Foundation for Fundamental Research on Matter (FOM) with financial 
aid from the Nederlandse Organisatie voor Wetenschappelijk Onderzoek (NWO).

\end{document}